\documentclass[sn-mathphys-ay]{sn-jnl}

\usepackage{graphicx}%
\usepackage{multirow}%
\usepackage{amsmath,amssymb,amsfonts}%
\usepackage{amsthm}%
\usepackage{mathrsfs}%
\usepackage[title]{appendix}%
\usepackage{xcolor}%
\usepackage{textcomp}%
\usepackage{manyfoot}%
\usepackage{booktabs}%
\usepackage{algorithm}%
\usepackage{algorithmicx}%
\usepackage{algpseudocode}%
\usepackage{listings}%

\newcommand*{\raa}{\ensuremath{\mathrm{Ra}=500}}
\newcommand*{\rab}{\ensuremath{\mathrm{Ra}=750}}
\newcommand*{\Nu}{\ensuremath{\mathrm{Nu}}}
\newcommand*{\Nubar}{\ensuremath{\overline{\mathit{Nu}}}}
\newcommand*{\Nuref}{\ensuremath{\mathrm{Nu}_{\mathrm{ref}}}}
\newcommand*{\NE}{\ensuremath{\mathrm{N}_{\mathrm{E}}}}
\newcommand*{\Ra}{\ensuremath{\mathrm{Ra}}}

\newcommand{\bs}[1]{\boldsymbol{#1}}
\DeclareMathOperator*{\E}{\mathbb{E}}

\usepackage{etoolbox}   
\makeatletter
\patchcmd{\@maketitle}{\artauthors}{{\artauthors}}{}{}
\makeatother

\theoremstyle{thmstyleone}%
\theoremstyle{thmstyletwo}%

\theoremstyle{thmstylethree}%

\begin{document}

\title[Article Title]{Multi-agent reinforcement learning for the control of three-dimensional Rayleigh--Bénard convection}

\author*[1]{\fnm{Joel} \sur{Vasanth}}\email{jvasanth@kth.se}

\author[2]{\fnm{Jean} \sur{Rabault}}

\author[1]{\fnm{Francisco} \sur{Alcántara-Ávila}}

\author[3]{\fnm{Mikael} \sur{Mortensen}}

\author*[1]{\fnm{Ricardo} \sur{Vinuesa}}\email{rvinuesa@mech.kth.se}

\affil*[1]{\orgdiv{FLOW, Engineering Mechanics}, \orgname{KTH Royal Institute of Technology}, \orgaddress{\city{Stockholm},  \country{Sweden}}}

\affil[2]{Independent Researcher, \orgname{Oslo}, \country{Norway}}

\affil[3]{\orgdiv{Department of Mathematics}, \orgname{University of Oslo}, \orgaddress{\city{Oslo}, \country{Norway}}}

\abstract{Deep reinforcement learning (DRL) has found application in numerous use-cases pertaining to flow control. Multi-agent RL (MARL), a variant of DRL, has shown to be more effective than single-agent RL in controlling flows exhibiting locality and translational invariance. We present, for the first time, an implementation of MARL-based control of three-dimensional Rayleigh--Bénard convection (RBC). Control is executed by modifying the temperature distribution along the bottom wall divided into multiple control segments, each of which acts as an independent agent. Two regimes of RBC are considered at Rayleigh numbers $\mathrm{Ra}=500$ and $750$. Evaluation of the learned control policy reveals a reduction in convection intensity by $23.5\%$ and $8.7\%$ at $\mathrm{Ra}=500$ and $750$, respectively. The MARL controller converts irregularly shaped convective patterns to regular straight rolls with lower convection that resemble flow in a relatively more stable regime. We draw comparisons with proportional control at both $\mathrm{Ra}$ and show that MARL is able to outperform the proportional controller. The learned control strategy is complex, featuring different non-linear segment-wise actuator delays and actuation magnitudes. We also perform successful evaluations on a larger domain than used for training, demonstrating that the invariant property of MARL allows direct transfer of the learnt policy.}


\maketitle

\section{Introduction}\label{sec:Introduction}
Reinforcement learning (RL) is a data-driven approach to the problem of goal-directed sequential decision making. The RL framework involves an `agent', which is its decision-making component, that learns a mapping from states to actions, called the `policy', via a series of trial-and-error interactions with the environment. A reward signal, which is a scalar measure of performance of the environment in relation to the pre-defined goal, is fed back to the agent and is used in the optimization. The state $S_t$, action $A_t$ and reward $R_t$ form the three fundamental channels of communication between the agent and the environment. Each state is assigned a value $V(s)$ which is the expected total reward that can be obtained by that state in the long run, \textit{i.e.}, $V(s)=\E \left[ \Sigma_{k=0}^{\infty} \gamma^k R_{t+k+1} | s \right]$ \citep{sutton2018reinforcement}. The goal of the RL problem, then, is to maximise expected cumulative rewards or `returns'. In Deep RL (DRL), function approximators such as neural networks are used to parameterise either the policy, the value function, or both, which are updated during training via interaction with the environment in order to maximise the returns. Various solution methods have been developed to perform this optimisation, namely: (i) value-based methods such as Q-learning method \citep{van2016deep, mnih2013playing}, where the value function is approximated and actions are chosen based on the approximated value function, (ii) policy-based methods \citep{silver2014deterministic}, such as the policy gradient method, where the optimal policy is learned directly, and (iii) actor-critic methods such as the proximal policy optimisation (PPO) method \citep{schulman2017proximal}, where the critic network computes an advantage function to determine how good an action is compared to the average of all possible actions, given a state. Then, the actor network is updated based on the value of this advantage function. During training, interaction between the agent and the environment occurs on an episode basis, where an `episode' is a single trajectory of the environment for a given fixed duration. Training is performed through trial-and-error that involves a stochastic exploration of the environment's states, \textit{i.e.}, selecting actions other than those dictated by the latest updated policy, which can improve the estimation of the state values. After training, the agent is evaluated on the environment in a deterministic manner according to the optimal policy, without exploration. \par 

DRL is well adapted to perform control of complex, high-dimensional, nonlinear systems. A successful application that contributed to its recent popularity is in gaming, where neural network-based agents are trained to play and win computer games such as Atari \citep{mnih2013playing}, Go \citep{silver2018alphago} and Starcraft \citep{vinyals2019grandmaster}. DRL has also been applied to industrial control problems \citep{degrave2022magnetic}, biomechanical optimization problems \citep{VermaEtAl2018}, and recently, a number of classical active flow control (AFC) problems \citep{Guastoni2023deep, sonoda2022reinforcement, vignon2023marlrbc, font2024active}. Compared to traditional control methods that are used in AFC, such as state-space methods and adjoint methods \citep{chevalier2002adjoint}, DRL does not require information about the governing equations or insight into the full state of the system. Even though there are a number of other techniques that aim at performing similar tasks as DRL \citep{brunton2015closed,brunton2020machine,pino2023comparative}, DRL has been found to be particularly flexible and effective in the recent years. While the trial-and-error learning method used in DRL can make it more data-hungry and computationally expensive to train, once trained, predicting the next action from the learned approximation to the optimal policy is computationally cheap. This, in conjunction with the fact that DRL can be adapted to large state spaces, makes it a viable option for real-world control applications, such as AFC. \par 

As a consequence, DRL for AFC has been the focus of much work over the last few years. One of the first uses of DRL in AFC was for the stabilisation of the vortex street in the flow past a two-dimensional (2D) cylinder at a low Reynolds number ($Re$) of 100 \citep{rabault2019artificial,rabault2019accelerating}. Here, the Reynolds number is given as $Re=\tilde{U}\tilde{H}/\tilde{\nu}$, where the symbols $\tilde{U}$, $\tilde{H}$ and $\tilde{\nu}$ are the mean flow velocity, characteristic length scale of the system and kinematic viscosity respectively, and tilde $\tilde{\cdot}$ represents dimensional quantities. With two synthetic jets as the actuators, the drag was reduced by $8\%$ and the learned control strategy was found to reduce pressure drop across the cylinder by increasing the area of the separated region. Following this study, DRL was applied to different configurations of the 2D cylinder. In \citet{tang2020robust}, four synthetic jets were used as DRL controllers for a wider $Re$ range, $Re=100-400$, and a maximum drag reduction of $\approx 38\%$ was achieved. Drag reduction was observed for unseen values of $Re$ as well, showing the robustness of the DRL control process. The work by \citet{ren2021applying} extends these studies to nonlinear unsteady conditions at $Re=1000$, showing a reduction in drag by $30\%$ along with a reduction of turbulent fluctuations. \citet{varela2022deep} shows that a transition in the actuation strategy is obtained when the Reynolds number is further increased and turbulent stresses play a dominating role. \citet{fan2020reinforcement} and \citet{xu2020active} use a pair of smaller downstream rotating cylinders as actuators with a combined objective of drag reduction over the main cylinder and also to reduction of power loss due to cylinder rotation. More recently, an application to three-dimensional (3D) cylinders was studied in the $Re$ range of $100-400$ \citep{suarez2023afc}, where again, DRL successfully showed a reduction of drag of $\approx 10\%$ for $Re=400$. A number of studies have also considered questions related to the optimization of the full DRL control setup, including probe placement \citep{paris2023reinforcement,paris2021robust,yan2023stabilizing} and the use of time history from the state probes \citep{wang2024dynamic}. Moreover, applications related to vortex induced vibration \citep{ren2019active,chen2023deep,ren2024enhancing}, or hydrodynamic stealth \citep{ren2021bluff}, have also attracted interest. Aside from the cylinder, other applications include the control of chaotic flows such as the flow arising from the Kuramoto--Sivashinsky (KS) equation \citep{bucci2019control, xu2023reinforcement}, drag reduction in 3D channel flows \citep{Guastoni2023deep, sonoda2022reinforcement}, suppression of the instability of a one-dimensional (1D) falling liquid film \citep{belus2019exploiting}, suppression of 2D Rayleigh--Bénard convection (RBC) \citep{beintema2020controlling, vignon2023marlrbc} and other practical problems \citep{ren2019active, ren2021bluff, li2022reinforcement}. We refer the interested reader to some comprehensive reviews of DRL applied to flow control \citep{vinuesa2022flow, vignon2023review, rabault2020deep, garnier2021review} for more detail. This rapid development has been made possible largely thanks to both the availability of open source DRL frameworks that make it easy to deploy advanced DRL algorithms \citep{kuhnle2017tensorforce,TFAgents}, and open source computational fluid dynamics (CFD) tools \citep{alnaes2015fenics,jasak2007openfoam}. These are being increasingly merged into turn-key frameworks that make it easy to start applying DRL to CFD simulations \citep{wang2022drlinfluids}. \par 

In relation to the works described above, several associated challenges have been solved over the last few years. For example, \citet{rabault2019accelerating} have leveraged a multi-environment DRL framework where several numerical simulations run in parallel, each providing its own set of $S_t,\ A_t$ and $R_t$ to a common agent, therefore scaling up the number of experiences. With their approach, they showed that speedups in training are obtained proportional to the number of simultaneous environments used. The study by \citet{belus2019exploiting} introduced a multi-agent RL (MARL) framework for the control of flow systems with distributed inputs and outputs. MARL was since then repeatedly proved to be effective in problems where the dynamics of the system are translationally invariant in space and multiple actuators are required to control the system, \textit{i.e.}, the control dimensionality is large. With a single-agent RL (SARL) framework, this large dimensionality would result in the curse of dimensionality on the action dimension making learning impractical. In the MARL framework, each agent is defined by separate streams of states, actions and rewards, each unique to a single agent. Moreover, critically, all agents share a common parameterisation, \textit{i.e.}, the same neural network weights. Note that here, we consider one of several possible configurations of MARL. An alternative MARL configuration, is where each agent has different individual parameterisations, and they are used to learn policies for different sub-tasks which contribute to a common goal. We refer the interested reader to the details of several such MARL variants in \citet{marl-book}. In the present work, each agent observes a small subdomain of the full domain, which we define as a `pseudo-environment' (Sec. \ref{subsec:methodology_drl}), and is thus responsible for only a subset of the total number of control actions, thereby overcoming the curse of dimensionality. The use of a common agent parameterisation is justified by the fact that the governing equations (and thus the qualitative features of the flow dynamics) are invariant over all pseudo-environments, and experience can be shared among MARL agents. Following \citet{belus2019exploiting}, recent works that employ MARL and have demonstrated its effectiveness are \citet{Guastoni2023deep} for the control of 3D turbulent channels, \citet{suarez2023afc,suarez2024active} to reduce drag over a 3D cylinder, \citet{Novati2021, Bae2022} for turbulence closure modelling and \citet{vignon2023marlrbc} for the control of 2D RBC, to name a few. \par 

The present work is an extension to 3D of the previous application of MARL to RBC in 2D presented by \citet{vignon2023marlrbc}. The RBC phenomenon is a fundamental problem widely studied academically in the field of thermally driven flow instabilities \citep{Drazin_Reid_2004}, which is also found in a range of natural phenomena and industrial applications. In our configuration, the domain consists of a fluid layer of height $\tilde{H}$ bounded by two rigid walls, where the temperature at the bottom wall $\tilde{T}_H$ is maintained at a constant value higher than the top wall temperature $\tilde{T}_C$. This forms an adverse temperature gradient $\Delta \tilde{T}/\tilde{H}=\left( \tilde{T}_H-\tilde{T}_C\right)/\tilde{H}$, which at a certain large enough value, destabilises the system. The instability occurs due to competing driving and damping effects, the driving effect being the vertical adverse thermal gradient and the damping effect being viscosity \citep{bergedubois1984}. The stability is governed by a non-dimensional number, the Rayleigh number \Ra{}, which is the ratio of the time scales of heat transport by conduction and by convection. The Rayleigh number is expressed as $\Ra=\tilde{g}\tilde{\alpha_T} \Delta \tilde{T}\tilde{H}^3/(\tilde{\nu}\tilde{\kappa})$, where $\tilde{g}$ stands for the acceleration due to gravity, $\tilde{\alpha_T}$ is the thermal expansion coefficient and $\tilde{\kappa}$ is the thermal diffusivity. At a certain critical $\Ra=\Ra_c$, instability occurs via a supercritical bifurcation from the no-motion purely conductive state to the convective state \citep{get98}. At large $\Ra>\Ra_c$, the heat transfer from the bottom wall to the top wall is dominated by convection. Another useful dimensionless number is the Nusselt number \Nu{}, which is defined as the ratio of the combined convective and conductive heat flux to the heat flux due to conduction alone. The Nusselt number is given as $\Nu = 1+\sqrt{\Ra \Pr}\ \langle vT \rangle_{x,y,z}$, where $\Pr$ is the Prandtl number ($Pr=\tilde{\nu}/\tilde{\kappa}$), $v$ and $T$ are non-dimensional vertical velocity and temperature and $\langle \cdot \rangle_{x,y,z}$ represents averaging over the entire domain. The manner of non-dimensionalisation is detailed in Sec. \ref{subsec:methodology_rbc}. \par 

Several approaches have been used to suppress the magnitude of convection in RBC systems. Passive flow control techniques \citep{kel92, car14, swa18} have shown only minor improvements in delaying the onset of convection (\textit{i.e.}, increasing the $Ra_c$). Active flow control techniques have proven to be more successful. Several AFC techniques involve modulation of the bottom wall temperature distribution \citep{sin91, wan92, or03, how97a, rem07, tan93a}. \par 

The first application of DRL to RBC was presented by \citet{beintema2020controlling} in a 2D domain using the single-agent RL framework. They studied the control of RBC in a square domain confined by no-slip walls with the lateral walls being adiabatic, top wall being isothermal, and temperature control modulation imposed on the bottom wall. The $\Ra_c$ at which convection occurs was reduced as the DRL controller converted a single-cell configuration in the uncontrolled system to a vertically stacked double cell configuration, reducing the effective \Ra{}. In a following study by \citet{vignon2023marlrbc}, the MARL framework was used on a domain with periodic boundary conditions on the lateral walls, featuring two convection cells in the uncontrolled state. Given that MARL is advantageous for problems with distributed inputs and outputs for which invariants are observed, a superior performance compared to SARL was demonstrated, in terms of faster learning. MARL achieved a $22.7\%$ reduction in \Nu{}, employing a control strategy that coalesces the two convection cells into a single-cell state, thus reducing the convection intensity. By extension to \citet{beintema2020controlling} and \citet{vignon2023marlrbc}, in the present study, the RBC system is three-dimensional. Therefore, the actuators are now two-dimensional surfaces at the bottom of the 3D fluid domain, as opposed to one-dimensional segments at the bottom of a 2D domain. As a consequence, in order to match the necessity to cover a 2D surface with actuators,  the number of actuators is increased from 10 in \citet{vignon2023marlrbc} to $8\times 8=64$ in the present case. If we consider a single-agent RL approach, this would mean that the action produced by the DRL agent would be a vector of 64 values, leading to the curse of dimensionality \citep{belus2019exploiting, vignon2023marlrbc}. This further motivates the need for the MARL framework for the execution of control in 3D. As far as the authors' knowledge of the past literature goes, the present study is the first application of MARL to 3D RBC. \par 

In particular, the main contributions of the present work are the following:
\begin{itemize}
    \item We introduce the MARL framework for 3D RBC. In 3D, the control dimensionality is multiplicatively larger, therefore, the use of multiple local agents corresponding to a MARL setup is much preferred to the SARL setup. Two cases of Rayleigh numbers, \raa{} and \rab{}, are tested.
    \item We show that MARL is successful in reducing convection at both \Ra{} regimes. The MARL controller drives the system from an uncontrolled state with irregular unsteady convection rolls, to a controlled state with regularly arranged steady convection rolls. The regularly arranged convection rolls are similar to the shape of convection structures found at a lower \Ra{} than \raa{} and \rab{}. Given that the \Ra{} is representative of the regime of instability \citep{bergedubois1984}, with flows at lower \Ra{} being relatively less unstable than flows at higher \Ra{}, we make it evident that the MARL-controlled states at \raa{} and \rab{} resemble flow from a less unstable regime, \textit{i.e.}, a regime at lower \Ra{}.
    \item We demonstrate how MARL can exploit translational invariances to obtain transferable strategies, by testing a pre-trained agent trained at \raa{} on a domain of larger size. We show that MARL reduces convection on the larger-size domain as well, without any additional tuning or modification of the agent.
    \item We apply classical proportional control and compare the performance of the MARL controller against the proportional controller. The MARL controller is shown to provide a larger reduction in \Nu{} than the proportional controller in both \Ra{} cases studied.
\end{itemize}

The rest of the paper is organised as follows. Sec. \ref{sec:methodology} provides a detailed description and formulation of the RBC problem (Sec. \ref{subsec:methodology_rbc}), the numerical methods used for the CFD (Sec. \ref{subsec:methodology_numericalCFD}), and a description of the DRL control (Sec. \ref{subsec:methodology_drl}) and proportional control (Sec. \ref{subsec:methodology_proportional}) methods. In the results section, Sec. \ref{sec:results}, the uncontrolled states are described in Sec. \ref{subsec:results_baselines}, and results from the training are discussed in \ref{subsec:results_training}. The evaluation of the trained agent in deterministic mode is provided in Sec. \ref{subsec:results_deterministic} and results of deterministic evaluation in a larger domain are provided in Sec. \ref{subsec:results_larger_domain}. Proportional control is discussed in Sec. \ref{subsec:results_proportional}. We conclude the paper in Sec. \ref{sec:conclusion}. Multimedia (video) files are referred to in the caption of various figures in the text and information about these is provided in the `Supplementary Information' section in Appendix \ref{sec:appendix_supplementary_information}. We release the full code used in this study as open source materials, and the relevant information is provided in Appendix \ref{sec:appendix_code_release}.

\section{Formulation and Methodology}\label{sec:methodology}
\subsection{RBC Formulation}\label{subsec:methodology_rbc}
\begin{figure}[t]
\centering
\vspace{-0.8cm}
\includegraphics[width=0.75\textwidth]{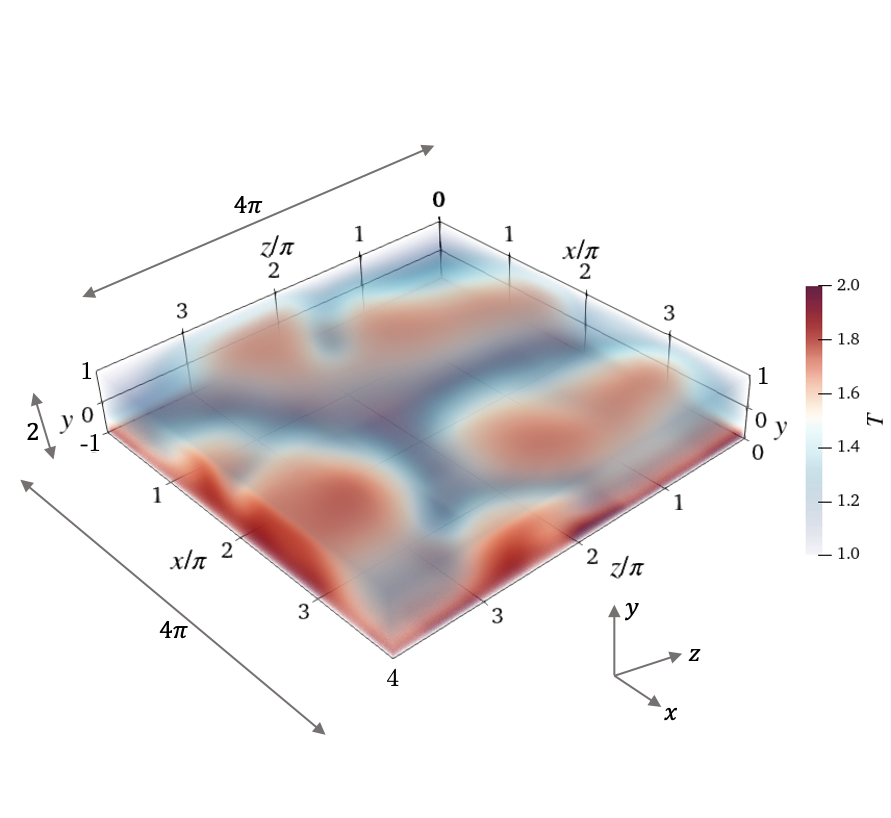}
\vspace{-0.8cm}
\caption{Illustration of the computational domain used in the simulation of the RBC system with its dimensions. Shown is the instantaneous temperature field along with convection rolls that form in the system as a result of the RB instability.}
\label{fig:domain}
\end{figure}

The formulation of the RBC problem closely follows that of \citet{vignon2023marlrbc}. The governing equations are the Navier--Stokes equations under the Bousinessq approximation, commonly adopted for convection problems. The Bousinessq approximation states that the density variation is only important in the gravitational term and neglects it in the rest of the equation. The continuity equation then reduces to the incompressible form, which further simplifies the viscous terms in the momentum equation. The viscosity is also assumed to be a constant. The equations are then made dimensionless by the domain height $\tilde{H}$, reference velocity $\tilde{U}_{\mathrm{ref}}=\sqrt{\tilde{g}\tilde{\alpha}_T\Delta \tilde{T}\tilde{H}}$ and the temperature difference $\Delta \tilde{T} = \tilde{T}_H - \tilde{T}_C$ between the bottom and top walls. Note that dimensional quantities are indicated with the tilde $\tilde{\cdot}$. The variable $\tilde{g}$ stands for the acceleration due to gravity, and $\tilde{\alpha}_T$ is the thermal expansion coefficient. Upon non-dimensionalisation, the coefficients can be refactored in terms of two non-dimensional constants, \textit{i.e.}, the Rayleigh number $\Ra=\tilde{g}\tilde{\alpha_T} \Delta \tilde{T}\tilde{H}^3/(\tilde{\nu}\tilde{\kappa})$ and Prandtl number $\Pr=\tilde{\nu}/\tilde{\kappa}$. The variable $\tilde{\nu}$ is the kinematic viscosity and $\tilde{\kappa}$ is the thermal diffusivity. Denoting the unit vectors in the $x$, $y$ and $z$ directions as $\mathbf{i}, \mathbf{j}$ and $\mathbf{k}$ respectively, the governing equations are \citep{pan18}:
\begin{subequations}
\label{eq:governing_equations}
\begin{equation}\label{eq:continuity}
\nabla \cdot \textbf{u} = 0,
\end{equation}
\begin{equation}\label{eq:momentum}
\frac{\partial \textbf{u}}{\partial t} + \left(\textbf{u}\cdot\nabla\right)\textbf{u} = -\nabla p + \sqrt{\frac{\Pr}{\Ra}}\nabla^2 \textbf{u} + T \mathbf{j}, 
\end{equation}
\begin{equation}\label{eq:temperature}
\frac{\partial T}{\partial t} + \textbf{u}\cdot\nabla T = \frac{1}{\sqrt{\mathrm{Ra} \Pr}}\nabla^2 T. 
\end{equation}
\end{subequations}
The velocity vector $\mathbf{u}(x,y,z,t)$ comprises of the three velocity components $\mathbf{u}=u\mathbf{i}+v\mathbf{j}+w\mathbf{k}$, and $T(x,y,z,t)$ is the temperature. The computational domain is cuboidal in shape with coordinates $x$ and $z$ representing horizontal directions and $y$ representing the vertical direction. The variable $t$ represents time. The length $L$ of the domain is measured along $x$ and $z$ and the height $H$ along $y$. The domain extends in the horizontal direction as $x,\ z\in[0,L]$ and in the vertical direction as $y\in[-1,1]$. Note that in this work, $L=4\pi$ and $H=2$. A schematic representation of the domain is shown in Fig. \ref{fig:domain}. To set up the temperature gradient, the top wall is at all times maintained at a spatially uniform temperature of 1, \textit{i.e.}, $T_C=T(x,y=1,z,t)=1$. The bottom wall is allowed to have a spatial distribution, \textit{i.e.}, $T_H(x,z,t)=T(x,y=-1,z,t)\neq \mathrm{constant}$, but we enforce that the spatial mean, \textit{i.e.}, $\langle T_H(x,z,t) \rangle_{x,z}=T_{H,0}$, is equal at all time to 2. The operator $\langle \cdot \rangle$ represents averaging in space. Note that $T_H$ is also spatially uniform during the uncontrolled simulations, \textit{i.e.}, $T_H(x,z,t)=2$, and it is only during the control simulations that the constant mean temperature constraint is enforced at the bottom wall. For the initial condition, a constant temperature gradient is provided as the temperature field, and quiescent flow for the velocity field. Periodic boundary conditions for velocity and temperature are provided at the lateral boundaries, while a no-slip boundary condition is prescribed at the top and bottom walls, along with the isothermal $T_C=1$ boundary at the top wall and constant $T_{H,0}=2$ at the bottom wall. \par 

A useful dimensionless number that can be used to quantify the magnitude of convection is the Nusselt number \Nu{}. The \Nu{} is defined as the ratio between the combined heat fluxes due to convection and conduction, and the heat flux due to conduction alone \citep{pan18}:
\begin{equation}\label{eq:nusselt_number}
    \Nu(t) = 1+\sqrt{\Ra \Pr}\ \langle vT \rangle_{x,y,z}.
\end{equation}

\subsection{Numerical Method for the CFD}\label{subsec:methodology_numericalCFD}
Following \citet{vignon2023marlrbc}, the governing equations~\eqref{eq:governing_equations} along with boundary conditions are implemented with a spectral Galerkin method using a technique developed by \citet{kim_moin_moser_1987} for direct numerical simulations of turbulent channel flows. In this method the pressure is eliminated and the 3D Navier--Stokes equations are basically reduced to the continuity equation, a fourth-order equation for the wall-normal velocity component and a second-order equation for the wall-normal vorticity component. Adding the energy equation, the four scalar equations that are solved are Eqs. \eqref{eq:continuity}, \eqref{eq:temperature}, as well as:
\begin{align}
    \frac{\partial \nabla^2 {v}}{\partial t} &= \frac{\partial^2 H_x}{\partial x \partial y} + \frac{\partial^2 H_z}{\partial z \partial y} - \frac{\partial^2 H_y}{\partial x^2} -\frac{\partial^2 H_y}{\partial z^2}  + \sqrt{\frac{\Pr}{\Ra}} \nabla^4 {v}  + \frac{\partial^2 T}{\partial x^2} + \frac{\partial^2 T}{\partial z^2}, \label{eq:momentumy}  \\
    \frac{\partial g}{\partial t} &= \sqrt{\frac{\Pr}{\Ra}} \nabla^2 g + \frac{\partial H_x}{\partial z} - \frac{\partial H_z}{\partial x}, \label{eq:gequation}
\end{align}
where $\bs{H} = (\bs{u} \cdot \nabla) \bs{u}$ is the convection vector and $g=(\nabla \times \bs{u}) \cdot \mathbf{j}$ is the normal vorticity component. The initial condition for the temperature is a linear gradient in the $y$-direction with the bottom and top wall temperatures maintained at 2 and 1 respectively. Equation \eqref{eq:momentumy} is implemented with the four boundary conditions $v(x, \pm 1, z, t) = v'(x, \pm 1, z, t) = 0$, where the first two are due to no slip, whereas the two latter follow from the continuity equation. Equation \eqref{eq:gequation} is derived by taking the curl of \eqref{eq:momentum}, and the $y$-component $g$ is subsequently solved with $g(x, \pm 1, z, t)=0$ (no slip). Periodic boundary conditions in the $x$ and $z$ directions are applied between the lateral boundaries, and this is implemented through the choice of the numerical elements and basis functions, as described below. \par 

The four scalar equations \eqref{eq:continuity}, \eqref{eq:temperature}, \eqref{eq:momentumy} and \eqref{eq:gequation} are implemented using a highly-accurate spectral Galerkin discretization in space \citep{shenbook} and a third-order implicit/explicit (IMEX) Runge-Kutta method \citep{ascher97} for the temporal integration. The Galerkin method makes use of tensor product basis functions constructed from Chebyshev polynomials for the wall-normal $y$ direction and Fourier exponentials for the periodic directions $x$ and $z$. The boundary conditions in both directions are built into the basis functions and as such enforced exactly. For the wall-normal direction this requires the use of composite Chebyshev polynomials \citep{shenbook}, whereas the Fourier exponentials for the $x$-direction are already periodic. Since the continuity equation cannot be used to find $u$ for Fourier wavenumber 0, we solve for this wavenumber the momentum equation in the $x$ direction. All other unknowns are closed through Eqs. \eqref{eq:continuity}, \eqref{eq:temperature}, \eqref{eq:momentumy} and \eqref{eq:gequation}. The convection terms $\bs{H}$ and $\bs{u} \cdot \nabla T$ are computed in physical space after expanding the number of collocation points by a factor of $3/2$ in order to avoid aliasing. For $\bs{H}$ we use the rotational form $\bs{H} = -\bs{u} \times (\nabla \times \bs{u})$, with the remaining $1/2 \nabla \bs{u} \cdot \bs{u}$ absorbed by the pressure, and for temperature we use the divergence form $\bs{u} \cdot \nabla T = \nabla \cdot \bs{u}T$. \par 

The code is implemented using the open-source spectral Galerkin framework `shenfun'~\citep{mortensen_joss}, where the equations can be automatically discretized through a high-level scripting language closely resembling the Mathematics. The Navier--Stokes solver has been verified by reproducing the growth of the most unstable eigenmode of the Orr--Sommerfeld equations over long time integrations. The Navier--Stokes and Rayleigh--Bénard solvers are distributed as part of the shenfun software, and there is a demonstration guide published in the documentation \citep{shenfun}, which also provides a much more detailed description of the numerical method. Links are provided in Appendix \ref{sec:appendix_code_release}, and we refer the reader interested in all the technicalities of the numerical solver implementation to the resources detailed therein. \par 

The observation probes are distributed over the domain as a uniform mesh. For purposes of illustration, probes over four randomly selected segments (to avoid overcrowding) are shown in black dots in Fig. \ref{fig:marl_setup}. There are $32 \times 8 \times 32$ total probe-mesh points in the $x$, $y$ and $z$ directions, respectively. The number of quadrature points in the solution approximation used by the spectral Galerkin solver is equal to the number of Galerkin modes, which is $32 \times 16 \times 32$. The resulting solution is spatially continuous, and the observations are then evaluated from these global, continuous spectral Galerkin functions at the uniform mesh probe locations. These observations are used in the DRL-based control methodology which we describe next.

\subsection{DRL Control Methodology}\label{subsec:methodology_drl}
The numerical simulations is solved with the methods described in the previous section from the constant temperature gradient initial condition up to a time instant when the time-averaged \Nu{}, \Nubar{}, becomes statistically constant and the RBC flow structures are fully-developed. By `fully-developed', we mean that the flow variables $u$, $v$, $w$ and $T$ (and by consequence, \Nu{}) are statistically stationary. Note that the overbar denotes averaging with respect to time. The \Nubar{} averaged over the last half of the total baseline simulation time is denoted as \Nuref{}, which is the reference \Nu{} used in the reward computation, as will be described later in this section. We refer to these uncontrolled initial simulations as the `baseline', since states at the final instants of these simulations serve as initial conditions for the control. For \raa{}, the time duration of the baseline is $400$ time units and for \rab{} it is $5000$. \par 

The control is applied to the RBC system as temperature actuations on the bottom wall boundary at $y=-1$. The bottom wall boundary is divided into $8\times 8$ square segments all of which are of equal size of $\pi/2 \times \pi/2$ in the $x$ and $z$ directions. The RBC system with the bottom wall segments together form the DRL environment. As mentioned in Sec. \ref{sec:Introduction}, DRL occurs via interactions between three fundamental channels of communication between the environment and the agent: the state $S_t$, action $A_t$ and reward $R_t$ at time $t$. As per the SARL framework, $S_t$ comprises all the physical quantities measured at observation probes distributed within the domain, $R_t$ is a measure based on the performance of the system, and $A_t$ are the set of outputs from the agent which are control actuations. In the case of the RBC system $S_t$ can, for example, be values of $[T,\ u,\ v,\ w]$ at the observation probes, $R_t$ can be a function of \Nu{} and $A_t$ the set of 64 temperature actuations $T_{\mathrm{act}}$ applied on each segment. \par 

\begin{figure}[t]
\centering
\includegraphics[width=1.0\textwidth]{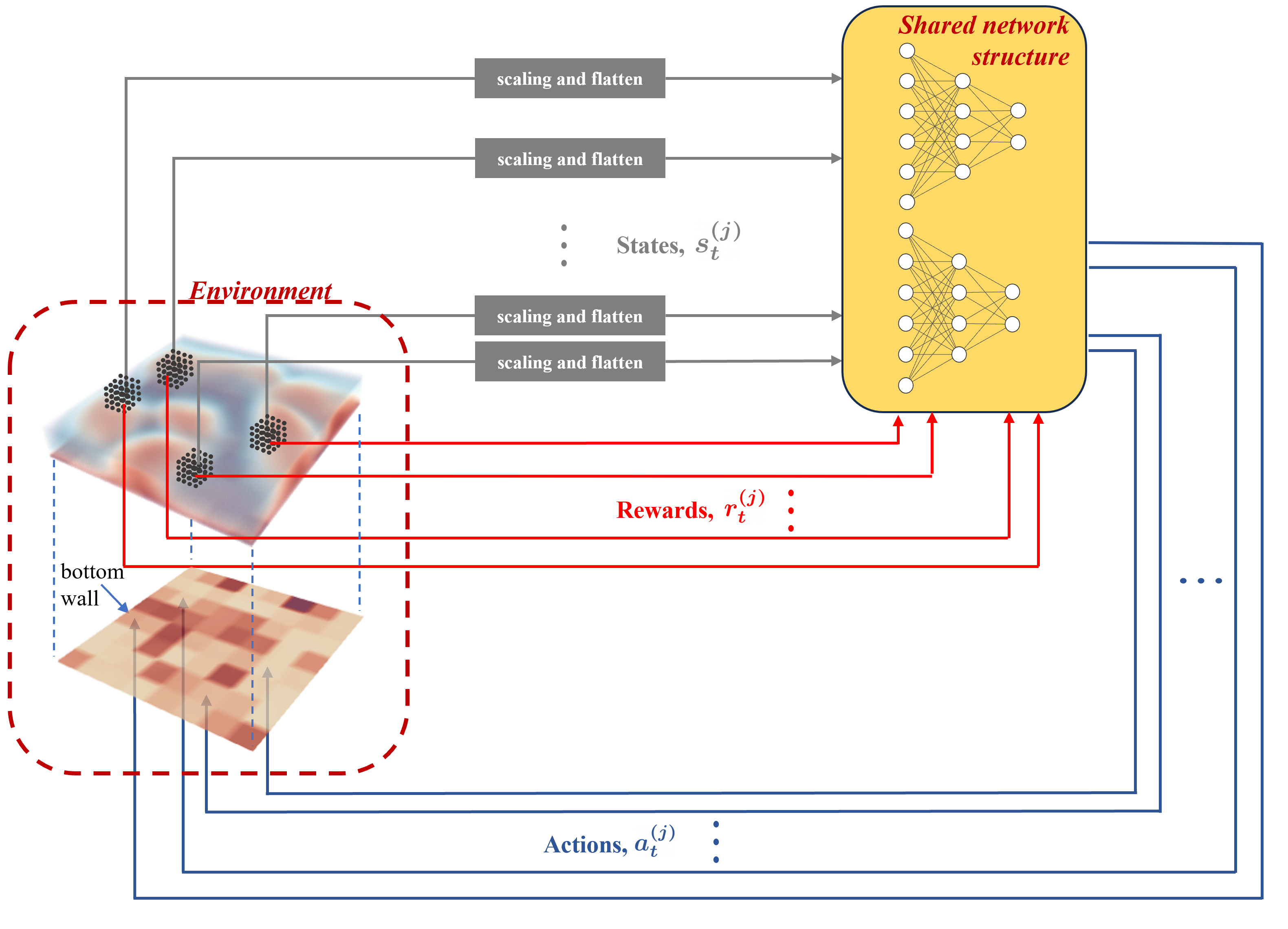}
\caption{An illustration of the MARL setup. The environment consists of the RBC system (top) with a projection of the bottom wall boundary (below) also shown. The bottom boundary is divided into $\mathrm{N}_{\mathrm{ag}}$ square segments where temperature control actuations are applied. Observation probes are distributed uniformly throughout the domain, but for illustration purposes are depicted with black dots in blocks of the domain directly above four pseudo-environments. Each pseudo-environment along with their streams of actions, states and rewards form multiple agents. The index $j$ represents the $j^{\mathrm{th}}$ agent, with $j=1,\dots, \mathrm{N}_{\mathrm{ag}}$. All agents share the same neural network parameterisation in the PPO algorithm. A schematic of two networks are shown representing the actor and critic parameterisation.}
\label{fig:marl_setup}
\end{figure}

In the following, we use a MARL setup as illustrated in Fig. \ref{fig:marl_setup}. In MARL, $\mathrm{N}_{\mathrm{ag}}$ multiple agents are each assigned to different local sections of the entire domain, which we refer to as `pseudo-environments'. We define a pseudo-environment as the block of the domain directly above each individual segment. Each agent is associated with its own stream of state, action and reward from its corresponding pseudo-environment which we denote as $s^{(j)}_t$, $a^{(j)}_t$ and $r^{(j)}_t$ respectively, $j=1,\cdots, \mathrm{N}_{\mathrm{ag}}$ being the agent index. Here, note the modification in notation from uppercase letters to lowercase letters: uppercase representing states, actions and rewards corresponding to the entire domain as per the SARL framework, and lowercase representing states, actions and rewards specific to each pseudo-environment as per the MARL framework. We represent segment-specific values with the superscript $j$ in parenthesis. Note that the $r^{(j)}_t$ is a `weighted' local reward, which is a combination of a `local' and `global' \Nu{} as described below. This is based on the fact that the RBC environment is invariant in the horizontal directions $x$ and $z$, so maximising a local reward can correspond to maximising the global reward in a cumulative manner. The state, action and reward tuples from each pseudo-environment and then from each timestep are batched together consecutively up to a size corresponding to the batch size, before an agent update is performed.

A key aspect of the MARL implementation that exploits translational invariance is that all the trainable weights in the agent parameterisation are shared, \textit{i.e.}, all agents, defined by their separate streams of state, action and reward, share the same neural network. This is an effective way to share experience across the domain and alleviate the curse of dimensionality over the control space dimension, as previously highlighted in \citep{belus2019exploiting,vignon2023marlrbc}. \par 

The states $s^{(j)}_t$ from each pseudo-environment are obtained from a set of $4\times 8\times 4 = 128$ observation probes distributed with even spacing in the $j^{\mathrm{th}}$ pseudo-environment. A representation of this is shown in Fig. \ref{fig:marl_setup} within the RBC domain in four randomly selected pseudo-environments. Four physical quantities: temperature and three velocity components $[T,\ u,\ v,\ w]$ are observed from each probe in each pseudo-environment. A scaling or shifting is performed to ensure that the state values are roughly between $-1$ and $1$ and not too small. The velocities are scaled as $u_s=1.5\times u$, $v_s=1.5\times v$ and $w_s=1.5\times w$ and the temperature is shifted as $T_s=T-1.8$, where subscript $s$ stands for `scaled' or `shifted'. The scaling factors used are based on the range of values observed in the baseline so that the scaled variables are roughly within the range $[0,1]$. These 512 quantities (4 variables measured at 128 observation probes) are flattened to form the local state vector $s_t^{(j)}=\left[T_s,\ u_s,\ v_s,\ w_s\right]^{(j)}$. 

The goal of any DRL algorithm is formulated as a maximisation problem, {\it i.e.}, to maximise the reward. In the MARL framework, the definition of reward for the $\mathrm{j}^{\mathrm{th}}$ pseudo-environment is slightly modified: it incorporates two \Nu{} values, one calculated within the $j^{\mathrm{th}}$ MARL pseudo-environment called the local Nusselt number $\Nu^{(j)}$, and one calculated over the entire domain called the global \Nu{}. The global \Nu{} is calculated using Eq. \eqref{eq:nusselt_number}. The local \Nu{}, $\Nu^{(j)}$, is given as: 
\begin{equation}\label{eq:local_nusselt_number}
    \Nu^{(j)}(t) = 1+\sqrt{\Ra \Pr}\ \langle vT \rangle_{x^{(j)},y,z^{(j)}},
\end{equation}
where $x^{(j)}$ and $z^{(j)}$ are the horizontal coordinate intervals corresponding to the $j^{\mathrm{th}}$ pseudo-environment, \textit{i.e.}, in the $j^{\mathrm{th}}$ pseudo-environment. The global \Nu{} is the part of the reward that accounts for the main goal of the agent, to minimise the convection over the entire channel. This incentivizes each pseudo-environment to improve the global flow state. The local \Nu{} provides information to the agent about the local effects of the actions on the flow. This allows for more reward granularity to the agent during training. Given that the RBC control goal is to minimise \Nu{}, we define the total reward for the $j^{\mathrm{th}}$ pseudo-environment to be a weighted reward $r^{(j)}_t$ proportional to the negative of a weighted sum of the global and local \Nu{}, defined as: 
\begin{equation}\label{eq:reward}
    r^{(j)}_t = \Nuref - \left[\beta \Nu^{(j)}(t) + (1-\beta)\Nu(t)\right],
\end{equation}
where \Nuref{} is the reference value equal to \Nubar{} with time-averaging performed over the last half of the total baseline simulation time and the factor $\beta=0.0015$ is a weighting factor determining the relative weight of the local and global \Nu{} contained in the reward. From Eq. \eqref{eq:reward}, a lower local or global \Nu{} results in a higher reward. The value of $\beta$ appears low, but this is due to the difference in geometric width used to compute the local and global \Nu{}'s. We choose $\beta$ such that the local \Nu{} accounts for about $10\%$ contribution to the total reward.

The output from the agents are actions $a_t^{(j)}$ in the range between $-1$ and $1$, each agent generating a single action for its corresponding control segment. These actions are then transformed into temperature actuations $T^{(j)}_{\mathrm{act}}$ applied to each bottom area using a shifting and normalisation process following:

\begin{align}
    a'^{(j)}_t &= a^{(j)}_t - \frac{1}{\mathrm{N_{ag}}}\sum_{j=1}^{\mathrm{N_{ag}}} a^{(j)}_t, \label{eq:actuations1} \\
    K &= \mathrm{max}\left( 1,\ \mathrm{max}_j\left| a'^{(j)}_t \right| \right), \label{eq:actuations2} \\
    T^{(j)}_{\mathrm{act}} &= T_{H,0} + \Lambda \frac{a'^{(j)}_t}{K}. \label{eq:actuations3}
\end{align}

In the above, each raw action is first subtracted by the mean over actions for all segments (\textit{i.e.}, for each MARL agent) to form action perturbations $a'^{(j)}_t$. The value $K$ is a factor to ensure that action perturbations larger than $1$ are scaled down before being converted into temperatures. The factor $\Lambda$ controls the magnitude of the temperature actuations. The above equations ensure that the mean temperature at the bottom boundary is kept at a constant value of $T_{H,0}=2$. This is the constant mean temperature constraint as described in Sec. \ref{subsec:methodology_rbc}. The RBC flow regime is characterised by the \Ra{} which depends on $T_{H,0}$, and thus changing $T_{H,0}$ would modify this regime. Thus, the constant mean temperature constraint is to ensure that despite the application of the control values, the regime of the instability remains unmodified and the system is controlled within this regime. \par

The states, rewards and actions described above are implemented within the framework of the `Tensorforce' library \citep{tensorforce}. Tensorforce contains several optimisation algorithms for reinforcement learning. In the present work, the widely used policy gradient method, proximal policy optimisation (PPO), is employed to train the agent. PPO is an actor-critic method where the actor approximates the policy distribution and the critic approximates the value function of the current state \citep{sutton2018reinforcement}. Both the policy and the critic are represented by neural networks. \par 

We define an episode as a simulation spanning 200 actuations, each actuation lasting for 15 time steps. At this point, we make a distinction in terminology between two types of episodes used in the MARL framework: `CFD episodes’ and `MARL episodes’. Both are simulations spanning 200 actuations, however, the difference lies in the part of the domain considered in the simulation. A CFD episode is a simulation of the entire domain involving 200 sets of $\mathrm{N_{ag}}$ actuations, one set for each action step. A MARL episode considers only the 200 actuations as applied locally by a single pseudo-environment. Thus, the simulation of a single CFD episode implies that $\mathrm{N_{ag}}$ MARL episodes are simulated. All results in this paper are plotted with respect to CFD episodes only. Therefore, hereafter in the text and figures, the usage of the term ‘episode’ refers to ‘CFD episode’, unless otherwise specified. A set of other training hyperparameters used in this study is provided in Table \ref{tab:parameters}. \par 

In the present work, 200 and 400 total CFD episodes are simulated for \raa{} and \rab{} respectively, and each actuation in any given episode is applied for a duration of 15 timesteps. In order to ensure that the agent is also trained on states from longer simulations, we ensure that $20\%$ of the episodes start from the end of the previous episode, and the remaining $80\%$ starts from the end state of the baseline. In this way, the system also explores trajectories spanning a longer time across successive episodes durations, and thus the agent observes more states, including states representative of the long-term behavior of the system. We use a batch size of 3 CFD episodes (\textit{i.e.}, $3\times \mathrm{N}_\mathrm{ag}=3\times 64=192$ MARL episodes). We found that with lower batch sizes, training is largely unstable. We explain this observation as follows. Neural networks learn best from data that is uncorrelated. Since data from adjacent MARL environments is strongly (spatially) correlated, a batch size large enough must be chosen such that independent trajectories can be sampled during training. On the other hand, a batch size too large would make training sluggish. Other agent hyperparameters and CFD parameters are provided in Table \ref{tab:parameters}. Training is performed with a discount factor of $\gamma=0.99$. For details of the algorithmic implementation of MARL, we refer the interested reader to \citet{vignon2023marlrbc}. We follow the same code implementation therein.

\subsection{Proportional Control Methodology}\label{subsec:methodology_proportional}
We also employ a proportional controller in order to compare the performance of the DRL-based controller described in the previous section, for \raa{} and \rab{}. For proportional control, actuations proportional to the temperature perturbations are applied on the bottom wall segments. The size of the bottom wall segments and number of segments are the same as described in Sec. \ref{subsec:methodology_drl}. In order to compute the magnitude of the temperature perturbations during instability, we require a reference temperature quantity from a stable state devoid of any RB instability. For this, we consider a low $\Ra=100$, corresponding to a stable regime that features a no-motion purely conductive state with a constant temperature gradient. The temperature at the mid-plane $y=0$ is measured, and we denote the temperature averaged over the entire mid-plane as $T_{\mathrm{ref,100}}=\langle T(x,y=0,z,t) \rangle_{x,z,t}$. We use $T_{\mathrm{ref,100}}$ as the reference temperature. Next, for \raa{} and \rab{}, at each time instant of the simulation, the temperature perturbation $T'$ is measured as
\begin{equation}
    T'(x,z,t)=T_{\mathrm{ref,100}} - T(x,y=0,z,t).
\end{equation}
Note that the value of $T'$ at the portion of the mid-plane directly above each control segment is used as the observation. For the $j^{\mathrm{th}}$ segment, the observation is $T'(x^{(j)},z^{(j)},t)$. A control action $T_p^{(j)}$ proportional to $-T'(x^{(j)},z^{(j)},t)$ is then applied on the $j^{\mathrm{th}}$ segment on the bottom wall as: 
\begin{equation}
    T_p^{(j)}=-K_p \langle T'(x^{(j)},z^{(j)},t) \ \rangle_{x^{(j)}, z^{(j)}}.
\end{equation} 

For the constant of proportionality $K_p$, a range of values is tested in both \raa{} and \rab{}. For \raa{}, the $K_p$ range is from $0.1$ to $1.0$ in increments of $0.1$. For \rab{}, the range of $K_p$ is $0.1$ to $2.0$ is increments of $0.1$. A wider range is used for \rab{} since it is more unstable than \raa{} and may require a larger amplitude of control. The response of the system for each individual $K_p$ is simulated with the initial condition being the end state of the baselines, until a controlled state is reached.

\begin{table}[t]
\caption{Parameters used in the present study.}\label{tab:parameters}
\begin{tabular}{@{}ll@{}}
\toprule
\textbf{Parameter} & \textbf{Value} \\
\midrule
\textbf{CFD} \\
Prandtl number, $\Pr$    & 0.7 \\ 
Time step, $\mathrm{\Delta t}$    & 0.1     \\
Domain size    & $4\pi\times 2\times 4\pi$     \\
Number of Galerkin modes & $32\times 16\times 32$     \\
Number of probes per pseudo-environment    & $4\times 8\times 4$   \\
\\
\textbf{DRL} \\ 
Number of CFD episodes for \raa{}    & 200   \\
Number of CFD episodes for \rab{}    & 400   \\
Actions per CFD episode    & 200   \\
Action duration    & 15 time steps    \\
Number of agents, $\mathrm{N_{ag}}$    & $64$   \\
State size    & 512   \\
Number of hidden layers (actor and critic networks)    & 2   \\
Number of units per hidden layer (actor and critic networks)    & 128   \\
Batch size   & 3 CFD episodes   \\
Reward weighting factor, $\beta$   & 0.0015 \\
Temperature actuation magnitude factor, $\Lambda$ & 0.9 \\
Discount factor, $\gamma$   & 0.99 \\
PPO clipping factor   & 0.2 \\
Learning Rate   & 0.001 \\
Entropy regularisation coefficient   & 0.01 \\
Optimizer   & Adam \\
\\
\textbf{Proportional Control} \\
Reference temperature, $T_{\mathrm{ref, 100}}$   & 1.5 \\
\botrule
\end{tabular}
\end{table}

\section{Results and Discussion}\label{sec:results}

\subsection{Baselines}\label{subsec:results_baselines}

\begin{figure}[t]
\centering
\includegraphics[width=0.45\textwidth]{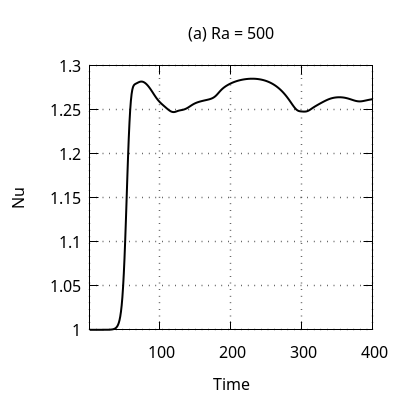}
\includegraphics[width=0.45\textwidth]{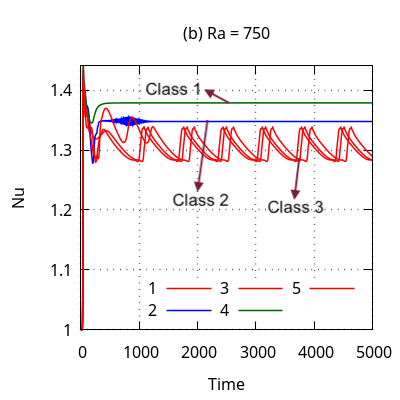}
\caption{Baseline simulations showing the evolution of \Nu{} with time $t$, from the purely conductive no-motion state ($\Nu{}=1$) to the onset of instability ($\Nu{}>1$) for (a) \raa{} and (b) five simulation runs labelled 1--5 for \rab{} categorized into three classes of baselines.}
\label{fig:BaselinesNu}
\end{figure}

Figure \ref{fig:BaselinesNu} shows the evolution of \Nu{} in the baseline simulations of each of the cases \raa{} and \rab{}. The value of the \Nu{} begins at 1.0, indicating the initial condition, \textit{i.e.}, the no-motion state where the heat transfer is due to conduction only. This state is unstable and the convective state sets in accompanied by a rise in \Nu{}. Then, \Nu{} stabilises at a given value or oscillates around a mean value, after which the baseline simulation is stopped. \par 

For \raa{}, five simulations were performed with the same parameters. The simulations all result in topologically similar solutions with similar \Nu{} values. We show a single solution in Fig. \ref{fig:BaselinesNu}(a). For \rab{}, similarly, five simulations of the baseline were performed with identical simulation parameters. However, in contrast to \raa{}, three different end-states are reached owing to the system chaoticity and stochasticity at \rab{}, some of which are oscillatory and some of which are steady. Since \Ra{} is higher, this state is more unstable and dynamically rich compared to \raa{}. We classify these baselines into three classes, class 1 to class 3, based on the magnitudes of \Nu{} after the transients have died away. The value of \Nubar{} is the obtained after time averaging \Nu{} over the last half of the baseline time. Class 1 corresponds to run 4, which shows a value of $\Nubar=1.379$, class 2 corresponds to run 2 with a lower steady \Nuref{} of $1.348$, and class 3 corresponds to simulations 1, 3 and 5 characterised by oscillatory \Nu{} values with $\Nubar=1.31$. For \rab{}, all of the baselines 1--5 are used to initialize the episodes while training the DRL agent, each episode starting from a randomly selected baseline.  The resulting \Nubar{} for \raa{} is $1.268$ and is used as \Nuref{} in Eq. \eqref{eq:reward}. For classes 1--3 of \rab{}, we use the average of all three \Nubar{} obtained at each of the three classes, obtaining a value of $\Nuref=1.346$.  \par

\begin{figure}[t]
\centering
\includegraphics[width=0.4\textwidth]{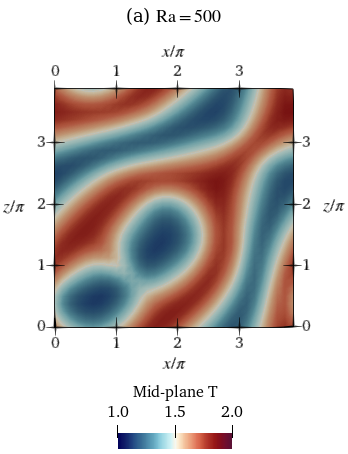}
\includegraphics[width=0.4\textwidth]{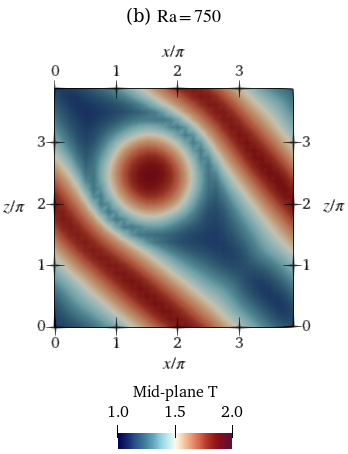}
\caption{Temperature fields of the baselines shown in the mid-plane cross-section ($y=0$) of the domain for (a) \raa{} at $t=400$ and (b) for \rab{} for class 3 (run 5) at time $t=5000$. Temperature fields of runs 1--4 of the other classes of baselines for \rab{} are shown in Appendix \ref{sec:appendix_convection_structure}. Video files are provided in Online Resource 1 and 2 for \raa{} and 750 respectively (see Appendix \ref{sec:appendix_supplementary_information}).}
\label{fig:BaselinesNu_topology}
\end{figure}

In Fig. \ref{fig:BaselinesNu_topology}, we plot the shapes of the convection rolls at the final time step of the baselines. Although training for \rab{} is performed with all 5 baseline runs, for brevity, Fig. \ref{fig:BaselinesNu_topology}(b) shows the topology for class 3 (run 5) only, since this is the baseline from which the deterministic agent evaluation and proportional control cases are simulated. The topologies of the remaining baseline runs for \rab{} are shown in Fig. \ref{fig:appendixA-ra750baselines} of Appendix \ref{sec:appendix_convection_structure}. At the warmest regions, the $v$ velocities are positive and maximum indicating upward flow, and in the coolest regions, $v$ is negative and minimum indicating flow towards the bottom wall. We define the length scale of the convection rolls as the distance between the warm and cold regions adjacent to each other at points where the temperatures are the maximum and minimum, respectively (\textit{i.e.}, $v$ is maximum and minimum). In both cases, the convection rolls have a length scale of approximately $\pi$, an observation in line with results of linear stability analysis \citep{bergedubois1984}. We perform a convergence analysis with different simulation time steps $\mathrm{\Delta t}$ to reinforce this result, which further justifies our choice of $\mathrm{\Delta t}=0.1$. The analysis shows that simulations with $\mathrm{\Delta t}\leq0.5$ are reasonable choices of time steps to obtain solutions with length scales of convection rolls that agree with linear stability analysis. Details of the convergence analysis are provided in Appendix \ref{sec:appendix_convergence}. \par

Note that videos of the temperature field evolution for these baselines are provided in Online Resource 1 for \raa{} and Online Resource 2 for \rab{}. Visible from the fields, all baseline shapes for \raa{} and 750 are irregularly shaped, with no particular spatial pattern. Class 2 and class 3 baselines for \rab{} exhibit spatial oscillations about a mean shape.

\subsection{Training}\label{subsec:results_training}

\begin{figure}[t]
\centering
\includegraphics[width=0.6\textwidth]{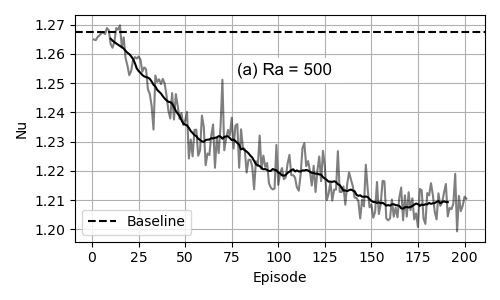}
\includegraphics[width=0.6\textwidth]{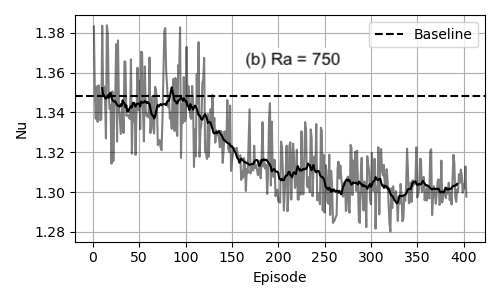}
\caption{Training curves for (a) \raa{} and (b) \rab{}. The episodes on the horizontal axis are CFD episodes. The dashed horizontal lines correspond to \Nuref{} from the baseline. Grey lines are episode-wise \Nu{} while black lines are for the moving average of \Nu{} over a window of 20 episodes.}
\label{fig:training_curves}
\end{figure}

Figure \ref{fig:training_curves} shows the training curves overlaid with their moving averages, with the horizontal axis being CFD episodes. The plotted values are the \Nu{} averaged over all time steps in each episode. The moving average curves are averaged over a moving window of 20 episodes. In both cases \raa{} and \rab{}, the mean \Nu{} reduces to values below the \Nuref{} in the baseline and then remains within a range of values. To measure the final \Nu{} after training, the plotted \Nu{} values are averaged over the last $\NE$ episodes. We denote this final \Nu{} after training as $\langle \Nu \rangle_{\NE}$, with $\langle \cdot \rangle_{\NE}$ representing averaging over the last \NE{} episodes. Thus, in each case, we can quantify the extent of reduction of convection by calculating the percentage decrease of \Nu{} from the reference value \Nuref{} using the following formula:

\begin{equation}\label{eq:nusselt_percent_reduction_training}
    \textrm{Percentage reduction of \Nu}=\frac{\Nuref-\langle \Nu \rangle_{\NE{}}}{\Nuref-1}\times 100\%.
\end{equation}

Note that the denominator is $\Nuref-1$ and not simply $\Nuref$. This is because the minimum possible \Nu{} in the RBC system is 1 (Eq. \eqref{eq:nusselt_number} with no convection, \textit{i.e.}, $v$=0) and so the percentage reduction calculation is offset by 1. For \raa{}, $\NE=50$, (\textit{i.e.}, averaging over episodes 150 to 200) and $\langle \Nu \rangle_{\NE{}}=1.209$. From the baseline $\Nuref=1.268$, this gives is a percentage \Nu{} reduction of $22\%$. \par 

\begin{figure}[t]
\centering
\includegraphics[width=0.9\textwidth]{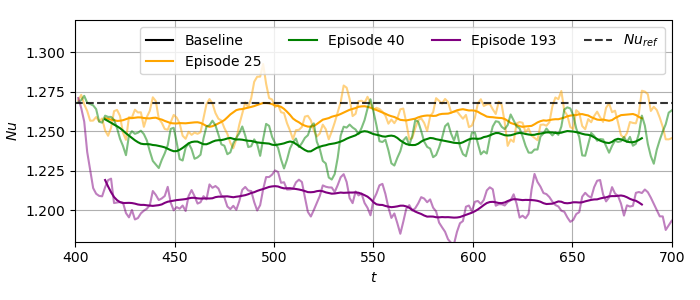}
\caption{Evolution of \Nu{} in three episodes during training at \raa. Transparent lines represent instantaneous values and bold opaque lines represent the moving-average over 30 time units.}
\label{fig:Ra500_nu3eps}
\end{figure}

In Fig. \ref{fig:Ra500_nu3eps}, we show the \Nu{} evolution during three episodes in different stages of the training. At episode 25, there is not much improvement compared to \Nuref{}. However, it is noticeable that the different actuations make the \Nu{} variation more noisy compared to the baseline. At episode 40 and more so at episode 193, a significant improvement is observed. In both cases, the \Nu{} trajectories first drop to a low value between $t=400-450$, and then remain low. In episode 193, the drop to a low \Nu{} occurs sooner ($t=400-410$) and to a much lower value compared to episode 40 ($t=400-450$). This shows that as the training proceeds, the agent learns a more optimal combination of actions to reduce convection more rapidly in the first few timesteps of control, and ensures that the system sustains this controlled state. \par 

In the training curve at \rab{} in Fig. \ref{fig:training_curves}(b), similar to \raa{}, the DRL agent reduces the \Nu{}, showing that a control strategy is learned. However, given that the \rab{} case is more unstable than \raa{} and five baselines are used during training, the phase space to be explored by the agent is much larger than for \raa{}, and so learning requires more steps, \textit{i.e.}, a larger number of episodes are required before a significant drop in \Nu{} is observed. The \Nu{} begins to drop at episode 100, compared to episode $\approx 20$ for \raa. After a gradual decrease in \Nu{} from episode 100--250, it saturates after episode 250. Choosing $\NE=150$,  $\langle \Nu \rangle_{\NE{}}=1.302$ giving a percentage \Nu{} reduction of $12.7\%$. \par

\begin{figure}[t]
\centering
\includegraphics[width=0.9\textwidth]{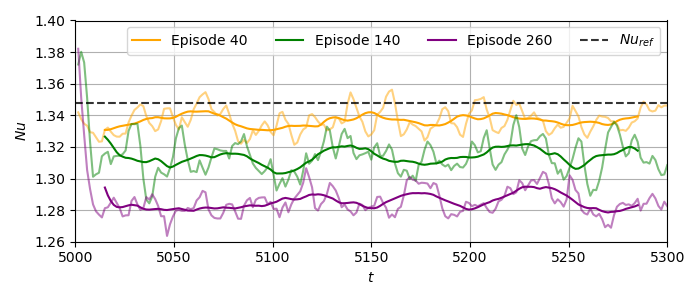}
\caption{Evolution of \Nu{} in three episodes during training at \rab.  Transparent lines represent instantaneous values and bold opaque lines represent the moving-average over 30 time units.}
\label{fig:Ra750_nu3eps}
\end{figure}

In Fig. \ref{fig:Ra750_nu3eps}, we plot the \Nu{} evolution during three episodes during the training at \rab{}. Similar to the \raa{} case, the progress to a more effective control policy is observed for higher episodes. Also similar to \raa{}, during initial times $t<5020$, the drop in \Nu{} from its initial value is quicker (\textit{i.e.}, a more negative slope) in later episodes compared to earlier ones. \par

In order to understand the nature of the control policy learned by the agent, we need to investigate the deterministic evaluations of the trained agent, which we show in the next section.


\subsection{Deterministic Evaluation}\label{subsec:results_deterministic}
In this section, we get an insight into the control method learned by the agent. In each case \raa{} and \rab{}, we load the agent saved at the final episodes of the training, \textit{i.e.}, episode 200 for \raa{} and episode 400 for \rab{}. The loaded agent is used in the evaluation mode, by choosing the most probable action for each observed state in a deterministic manner, in contrast to the training mode, when the agent explored the phase space by using exploration on the actions. In order to compute the percentage reduction in \Nu{} during the deterministic runs, we use the following formula: 
\begin{equation}\label{eq:nusselt_percent_reduction_deterministic}
    \textrm{Percentage reduction of \Nu}=\frac{\Nuref-\Nu_C}{\Nuref-1}\times 100\%,
\end{equation}
where $\Nu_C=\Nubar^{\Delta t_C}$ is the \Nu{} of the controlled state calculated by averaging the \Nu{} over some portion ($\Delta t_C$) of the time during which control is executed. In order to justify the offset by 1 in the denominator, the same reasoning is used as in Eq. \eqref{eq:nusselt_percent_reduction_training}.
\begin{figure}[t]
\centering
\includegraphics[width=0.45\textwidth]{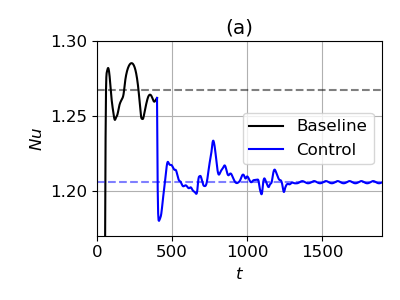} \\
\includegraphics[width=0.47\textwidth]{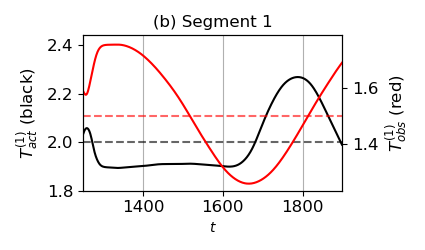}
\includegraphics[width=0.47\textwidth]{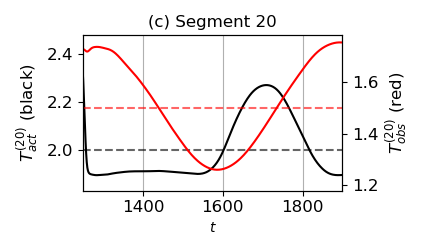}
\includegraphics[width=0.47\textwidth]{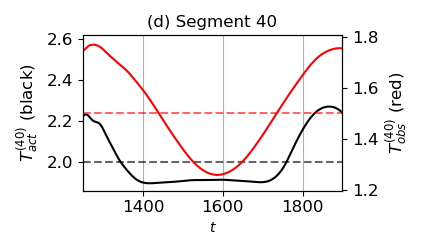}
\includegraphics[width=0.47\textwidth]{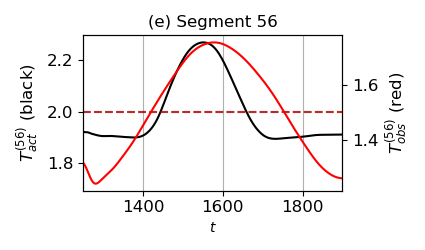}
\caption{(a) \Nu{} curve during the deterministic evaluation of the agent at \raa{}, plotted along with the baseline. The broken lines represent $\Nuref$ (black, top) from the baseline, and $\Nu_C$ at the controlled state (blue, below). (b)-(e) Evolution of control temperatures $T_{\mathrm{act}}^{(j)}$ (left axis, black lines) and temperature observation at the mid-plane $T_{\mathrm{obs}}^{(j)}$ (right axis, red lines) in randomly selected control segments from $t=1250-1900$. The horizontal broken lines represent the values of $T_{H,0}=2$ (left axis, black) and $1.5$ (right axis, red) respectively.}
\label{fig:Ra500_det}
\end{figure}

\subsubsection{\raa{}}
First, we look at the evaluation for \raa{} in Fig. \ref{fig:Ra500_det}. Control is executed from $t=400$. Within the first few timesteps, the \Nu{} drops to a low value. Following a period of transients, the system stabilises into a final controlled state from $t=1300-1900$. Thus, $\Delta t_C=600$ and following Eq. \eqref{eq:nusselt_percent_reduction_deterministic}, $\Nu_C=1.205$. This corresponds to a \Nu{} reduction of $23.5\%$. \par 

Next, we show the actions alongside the observations in a few segments in Fig. \ref{fig:Ra500_det} (b)-(e), to get an idea of the learned policy. Here, segments 1, 20, 40 and 56 are shown for time instants during the final controlled state. In each segment, we compare control temperatures $T_{\mathrm{act}}^{(j)}$ (Eq. \eqref{eq:actuations3}) with temperature observations $T_{\mathrm{obs}}^{(j)}$, which are temperatures in the mid-plane $y=0$ in the $j^{\mathrm{th}}$ pseudo-environment averaged in the horizontal directions, \textit{i.e.}, $T_{\mathrm{obs}}^{(j)}=\langle T(x^{(j)},y=0,z^{(j)},t) \rangle_{x^{(j)}, z^{(j)}}$. We also plot in dashed lines the mid-plane temperature at a stable state if there was no convection, which equals $1.5$. Noticeably, the control strategy is non-trivial and there exists a complex non-linear relation between the averaged temperature above a segment and the bottom wall temperature chosen by the DRL controller. In particular, we observe:

\begin{itemize}
    \item \textbf{Control actuations disproportionate to observations}: although $T_{\mathrm{obs}}^{(j)}$ are roughly sinusoidal with time, the variation of $T_{\mathrm{act}}^{(j)}$ does not follow the same pattern, but is instead composed of extended periods of actuations below $T_{H,0}=2$ and short peaks of warmer actuations above $T_{H,0}$. Also, in segments 1 and 20, as $T_{\mathrm{obs}}$ increases above $1.5$, $T_{\mathrm{act}}$ decreases, similar to opposition control. However, $T_{\mathrm{act}}$ remains at a low value even when $T_{\mathrm{obs}}$ begins to drop. In segments 40 and 56, $T_{\mathrm{obs}}$ and $T_{\mathrm{act}}$ have the same variation pattern: when $T_{\mathrm{obs}}$ decreases below (increases above) the mean, then $T_{\mathrm{act}}$ increases above its mean as well. Thus, the learned control strategy features actuations that are disproportionate to the observations in some segments. This is distinct from an opposition control strategy, where the actuations follow a pattern of variation opposite to the observations, uniformly across segments.
    \item \textbf{Differences in actuator delays per segment}: It is also visible from segments 1 and 20 that the $T_{\mathrm{act}}$ begins to increase after a short delay in $T_{\mathrm{obs}}$ increasing. This duration of delay is different across different segments. In the case of opposition control with delays, $T_{\mathrm{act}}^{(j)}$ would follow the exact opposite pattern of variation as $T_{\mathrm{obs}}^{(j)}$ across segments, and there would be no difference in delays from segment to segment. Moreover in opposition control, the delays in each segment are required to be tuned before the optimal control state is achieved, while in DRL control, these delays are implicitly learned in the control policy. 
\end{itemize}

\begin{figure}[t]
\centering
\includegraphics[width=0.8\textwidth]{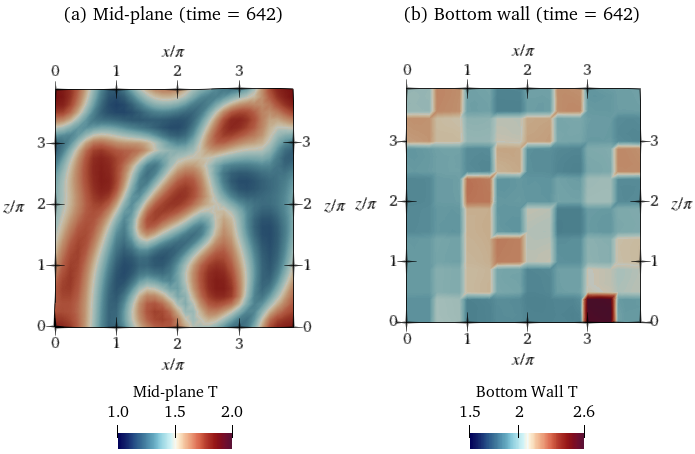}
\includegraphics[width=0.8\textwidth]{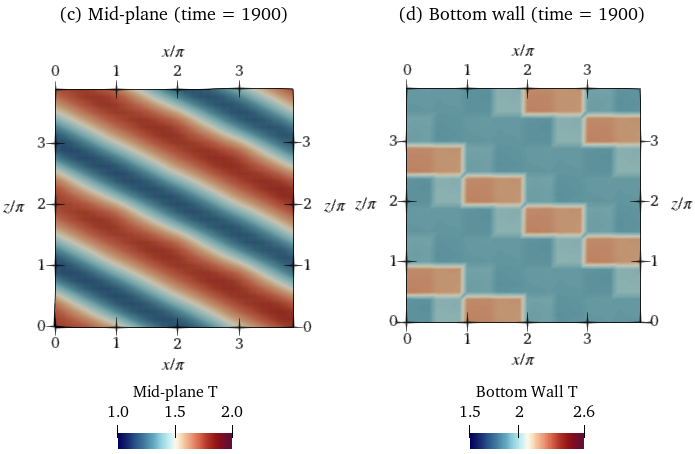}
\caption{Deterministic evaluation of the agent at \raa{}: temperature fields during (a)-(b) the transient phase at $t = 642$ and (c)-(d) at the final controlled state at $t = 1900$, shown at two cross-sections - the mid-plane of the domain (a) and (c) and the bottom wall showing the corresponding temperature actuations (b) and (d). Video provided in Online Resource 3.}
\label{fig:Ra500det_field}
\end{figure}

Next, we show in Fig. \ref{fig:Ra500det_field} how the flow structure corresponding to \raa{} is modified by the DRL control compared with that of the baseline (previously shown in Fig. \ref{fig:BaselinesNu_topology}). During the transient phase ($t=400-1300$), the DRL controller breaks down the convection rolls in the baseline state into smaller localised blobs (Fig. \ref{fig:Ra500det_field}(a)). Then, in the final controlled state beyond $t=1300$, the structure is rearranged to non-intersecting diagonally running convection rolls (Fig. \ref{fig:Ra500det_field}(c)). A sequence of spatial actuation patterns (two of which are shown in Fig. \ref{fig:Ra500det_field}(b) and Fig. \ref{fig:Ra500det_field}(d)) is executed by the agent in order to modify the flow topology and bring the system to this controlled state. This sequence of actuations is a feature of the learned control policy. \par 

\begin{figure}
    \centering
    \raisebox{0.25\height}{\includegraphics[width=0.5\textwidth]{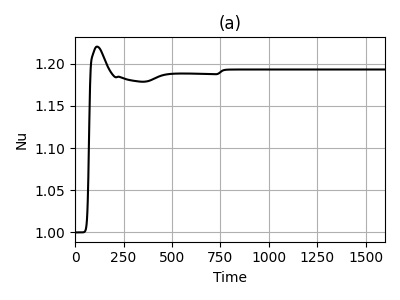}}
    \includegraphics[width=0.4\textwidth]{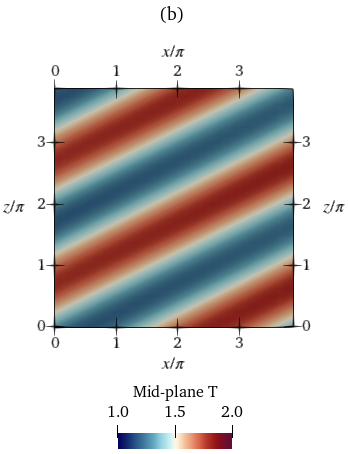}
    \caption{Evolution of \Nu{} (left) and temperature field at the mid-plane (right) from the baseline simulation at $\Ra=400$. The temperature field is shown corresponding to a time instant of 1500 (video in Online Resource 4).}
    \label{fig:Ra400}
\end{figure}

The regular convection structure of the controlled state is representative of RBC at a value of \Ra{} lower than \raa{}. To demonstrate this, we perform an additional baseline simulation at $\mathrm{Ra}=400$ and plot its \Nu{} and temperature field in in Fig. \ref{fig:Ra400}. $\mathrm{Ra}=400$ corresponds to a less unstable regime, featuring a smaller convection amplitude compared to \raa{}, as evident from Fig. \ref{fig:Ra400}(a). Further, the temperature field at instability (Fig. \ref{fig:Ra400}(b)) shows the regular pattern of diagonal convection rolls. It can thus be stated that the DRL controller modifies the trajectory of the system so that it resembles states observed in more stable regimes.

\subsubsection{\rab{}}
\begin{figure}[t]
\centering
\includegraphics[width=0.5\textwidth]{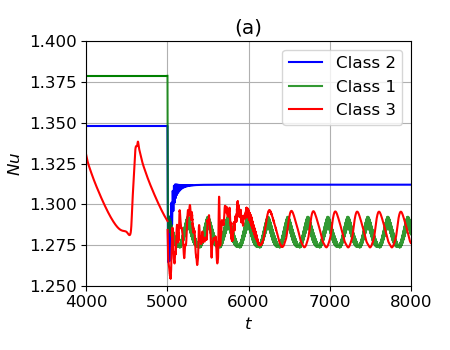} \\ 
\includegraphics[width=0.47\textwidth]{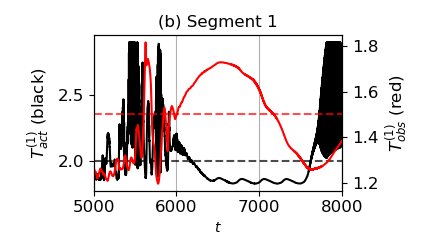}
\includegraphics[width=0.47\textwidth]{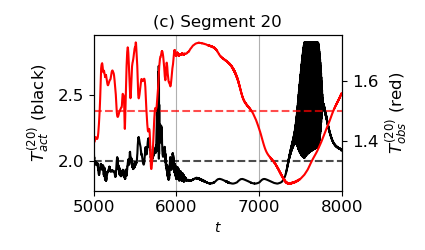}
\includegraphics[width=0.47\textwidth]{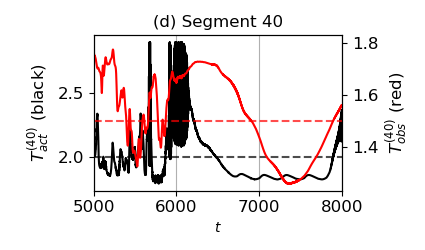}
\includegraphics[width=0.47\textwidth]{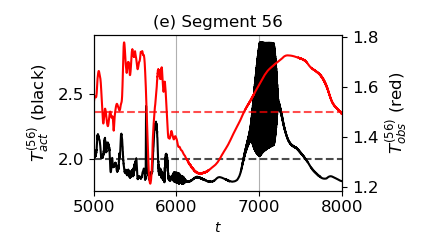}
\caption{(a) Evolution of \Nu{} during the deterministic evaluation of the agent at \rab{}, plotted for all three classes of baselines. Control begins at  $t=5000$. (b)--(e) Variation of temperature observation at the mid-plane $T_{\mathrm{obs}}^{(j)}$ (right axis) and control actuation $T_{\mathrm{act}}^{(j)}$ (left axis) in randomly selected control segments.}
\label{fig:Ra750_det}
\end{figure}

We now study the deterministic runs at \rab{}. We perform evaluations on all the three baseline classes as described in Sec. \ref{subsec:results_baselines}, using the latest saved agent from the training of \rab{}. The results are plotted in Fig. \ref{fig:Ra750_det}. In Fig. \ref{fig:Ra750_det}(a), we see that in all three classes, the agent reduces the \Nu{} to a lower value from the respective baselines. For classes 1 and 2, a controlled state is reached within approximately $200$ time units. For class 3, a controlled state is reached within $1000$ time units, \textit{i.e.} at time $6000$, after a period of transients. In order to calculate the percentage reduction in \Nu{}, the formula in Eq. \eqref{eq:nusselt_percent_reduction_deterministic} is used with the \Nuref{} values being the \Nu{} averaged over the last half of the baselines for each of the classes. As mentioned in Sec. \ref{subsec:results_baselines}, this is $1.379$, $1.348$ and $1.310$ respectively for classes 1, 2 and 3. From Fig. \ref{fig:Ra750_det}(a) we calculate the $\Nu_C$ reached in each class as 1.312, 1.281 and 1.283 respectively.  Therefore, for classes 1, 2 and 3, the agent achieves a percentage of \Nu{} reduction of $17.7\%$, $19.3\%$ and $8.7\%$, respectively. Given that the \rab{} agent is trained on five baselines, the ability to control all classes of baselines (a wider range of states) is an indicator of the robustness of the DRL agent.

To better understand the learned policy, we observe the plots in \ref{fig:Ra750_det}(c)--(e), where four segments are chosen at random to analyse the action chosen based on the observed states. Opposition control-like relationships are visible in segments 1 and 20, where from $t=6000-7000$, high temperature observations at the mid-plane cause the corresponding actuations to be lower than $T_{H,0}=2$. However, in segments 40 and 56, an opposite trend is visible. In segment 40 at $t=6500$, a decrease in $T_{\mathrm{obs}}$ is accompanied by a decrease in the actuation temperatures as well, and in segment 56 at $t=6500$, an increase in $T_{\mathrm{obs}}$ is accompanied by an increase in $T_{\mathrm{act}}$. Similar to \raa{}, the delays between the increase/decrease of $T_{\mathrm{obs}}$ and increase/decrease of $T_{\mathrm{act}}$ are segment-dependent and not consistent across segments, which differentiates the agent-learned control policy from a simple opposition control policy. The qualitative results of the learned control strategy are similar to that observed in \raa{}. The fact that such differences in the policies are observed, while the MARL setup implies that the DRL agent is shared across segments, indicates that complex non-linear strategies that depend on the state in a complex fashion have been chosen by the agent. \par 

\begin{figure}
    \centering
    \includegraphics[width=0.8\textwidth]{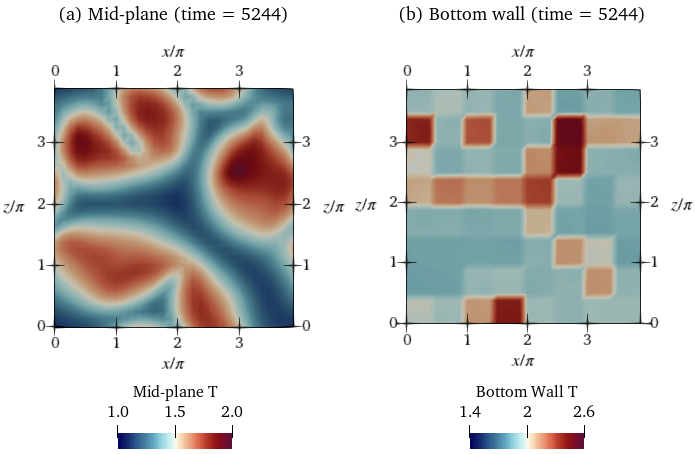}
    \includegraphics[width=0.8\textwidth]{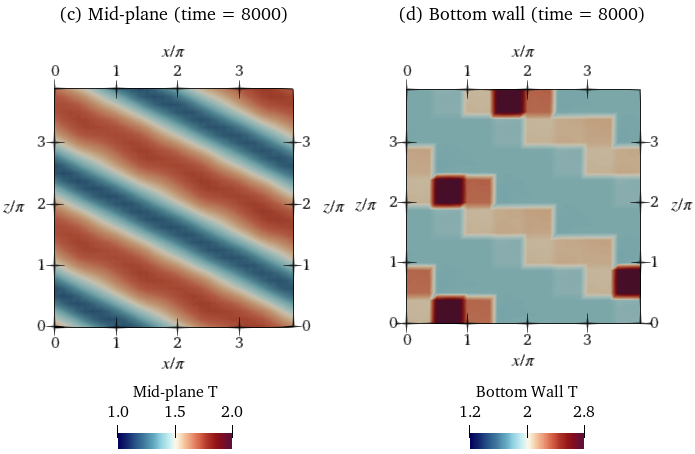}
    \caption{Deterministic evaluation of the agent at \rab{} for the class 3 baseline: shown are temperature fields during (a)--(b) the transient phase at $t = 5244$ and (c)--(d) at the final controlled state at $t = 8000$, at two cross-sections - the mid-plane of the domain (a) and (c) and on the bottom wall showing the corresponding temperature actuations (b) and (d). Video in Online Resource 5.}
    \label{fig:Ra750det_fields}
\end{figure}

Next, we plot the temperature fields of \rab{} during control in Fig. \ref{fig:Ra750det_fields} for the class 3 baseline. As is visible from the plot, the agent breaks down the convection cells in the baseline (previously shown in Fig. \ref{fig:BaselinesNu_topology}(b)) into smaller cells, reshaping the structure of the convection to straight rolls running diagonally across the domain. The final controlled state achieved by the agent resembles a convection pattern at a lower, less unstable $\Ra=400$ (Fig. \ref{fig:Ra400}).\par

\subsection{Proportional Control}\label{subsec:results_proportional}

\begin{figure}
    \centering
    \includegraphics[width=0.45\textwidth]{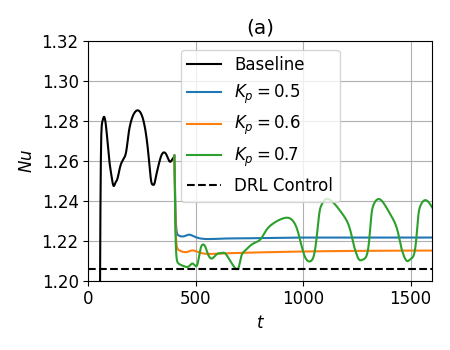}
    \includegraphics[width=0.48\textwidth]{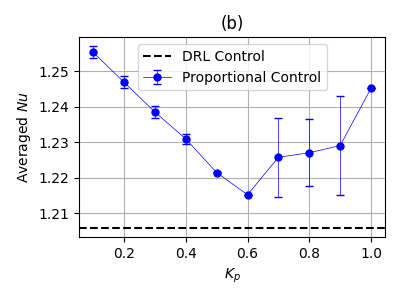}
    \caption{(a) Time evolution of \Nu{} with proportional control, shown for three different $K_p$ at \raa{}. (b) Time-averaged \Nu{}, averaged over the last half of the total control time duration, shown as a function of $K_p$. The error bars indicate the standard deviation of the \Nu{} from this averaged value, an indicator of the fluctuation amplitude.}
    \label{fig:Ra500prop_Nu}
\end{figure}

\begin{figure}
    \centering
    \includegraphics[width=0.48\textwidth]{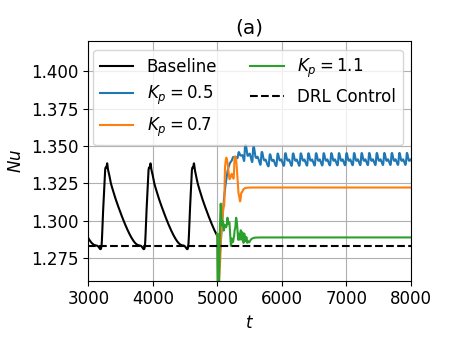}
    \includegraphics[width=0.5\textwidth]{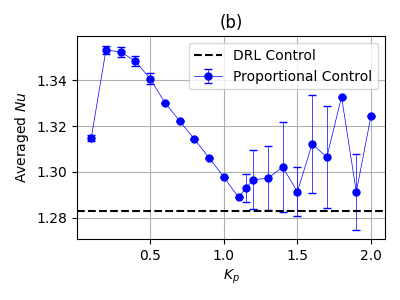}
    \caption{(a) Time evolution of \Nu{} during proportional control, shown for three different $K_p$ at \rab{}. (b) Time-averaged \Nu{}, averaged over the last half of the total control time duration, shown as a function of $K_p$. The error bars indicate the standard deviation of the \Nu{} from this averaged value, an indicator of the fluctuation amplitude. The class 3 baseline is used as the starting field.}
    \label{fig:Ra750prop_Nu}
\end{figure}

We compare the control policy learned by the DRL agent with a proportional controller applied to the Rayleigh--Bénard system as described in Sec. \ref{subsec:methodology_proportional}. In Figs. \ref{fig:Ra500prop_Nu} and \ref{fig:Ra750prop_Nu}, we show the response of the system to proportional control at \raa{} and \rab{}, respectively. In each case, the simulation is performed starting from the end state of the baseline. For \rab{}, the class 3 baseline is used, since this baseline was used in the deterministic run in Sec. \ref{subsec:results_deterministic}.  In Figs. \ref{fig:Ra500prop_Nu}(a) and \ref{fig:Ra750prop_Nu}(a), we only show the \Nu{} curves corresponding to the $K_p$ values with the best control performance. In Figs. \ref{fig:Ra500prop_Nu}(b) and \ref{fig:Ra750prop_Nu}(b), the \Nubar{} averaged over the last half of the total control time duration is plotted across all $K_p$. In both Figs. \ref{fig:Ra500prop_Nu} and \ref{fig:Ra750prop_Nu}, the mean \Nu{} of the DRL-controlled state, $\Nu_C$ is also shown for comparison. \par 

The reduction in \Nu{} for proportional control is computed using Eq. \eqref{eq:nusselt_percent_reduction_deterministic}. In Fig. \ref{fig:Ra500prop_Nu}, $K_p=0.6$ yields a controlled state with $\Nu_C=1.215$, corresponding to a percent reduction in \Nu{} of $19.8\%$. This is calculated from the \Nuref{} for \raa{} equal to 1.268. A drop in the averaged \Nu{} is observed as the $K_p$ is increased up to $0.6$, with either constant \Nu{}, or low-amplitude fluctuations. As $K_p$ is increased beyond $0.6$, overshooting occurs: the large control actuations cause the system to become further unstable, and the mean \Nu{} increases. Compared to DRL-based control in Fig. \ref{fig:Ra500_det}(a), the proportional control strategy is sub-optimal even at the best $K_p$, since DRL-based control reduces the value of \Nu{} to a lower value than the proportional control at $K_p=0.6$. This result suggests that with DRL, a control policy more sophisticated than a simple proportional control law is discovered. As discussed in Sec. \ref{subsec:results_training}, this corresponds to additional learned features of the control law such as a range of different actuator delays across segments and actuations disproportionate to temperature observations, corresponding to non-linear DRL control laws. \par 

\begin{figure}
    \centering
    \includegraphics[width=0.8\textwidth]{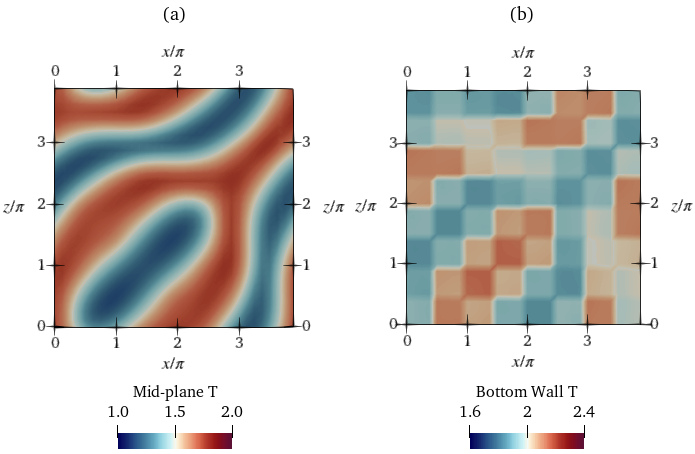}
    \caption{Temperature field at the (a) mid-plane and (b) bottom wall at \raa{}. The latter shows actuations across control segments for proportional control at $K_p=0.6$. Video shown in Online Resource 6.}
    \label{fig:Ra500prop_field}
\end{figure}

The temperature field at the final controlled state at $K_p=0.6$ for \raa{} is also plotted in Fig. \ref{fig:Ra500prop_field}. In contrast to the DRL-controlled end state, the convection rolls are not straight and parallel, and do not have uniform thickness, indicating that proportional control is not able to cause the same topological modifications obtained by the non-linear DRL controller.

\begin{figure}
    \centering
    \includegraphics[width=0.8\textwidth]{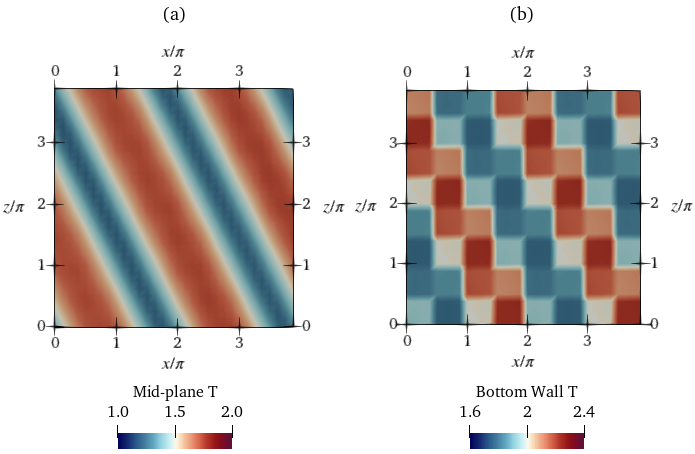}
    \caption{Temperature field at the (a) mid-plane and (b) bottom wall at \rab{}. The later shows actuations across control segments for proportional control at $K_p=1.1$. Video shown in Online Resource 7.}
    \label{fig:Ra750prop_field}
\end{figure}

At \rab{}, as is visible from Fig. \ref{fig:Ra750prop_Nu}(a), the $K_p$ corresponding to the largest \Nu{} reduction is $1.1$. Similar to \raa{}, the controlled-state mean \Nu{} reduces linearly with $K_p$ up to $K_p=1.1$ with a low-amplitude fluctuation about the mean \Nu{} at each $K_p$, and beyond $1.1$, the mean \Nu{} increases, \textit{i.e.}, overshooting occurs, and the system becomes unstable. A controlled state similar to the DRL-controlled state, with straight parallel cells running diagonally past the domain is observed in Fig. \ref{fig:Ra750prop_field}. The $\Nu_C$ at the controlled state achieved by the DRL agent is $1.283$, which is slightly lower than that of the proportional controller at $K_p=1.1$, which is $1.289$. The \Nuref{} of the class 3 baseline is $1.31$ (Sec. \ref{subsec:results_baselines}). The DRL controller achieves a percentage \Nu{} reduction of $8.7\%$, and with proportional control, a $6.8\%$ reduction is achieved. \par 

Although the \Nu{} reduction at \rab{} using DRL and proportional control are comparable, it is noteworthy that the agent at \rab{} is trained using three different baselines, and learns an effective control policy for all baselines during a single training run. The agent at \rab{} is thus more capable to generalise to a wider range of states, highlighting its robustness. In the case of a proportional controller, an efficient tuning of $K_p$ must be carried out for each of the three baseline classes until an optimum \Nu{} reduction is observed.

\subsection{Deterministic evaluation in a larger domain}\label{subsec:results_larger_domain}
One of the core advantages of MARL compared to SARL is its ability to reduce the state and action space dimensionality by exploiting spatial invariances in the system. This also allows to easily transfer a MARL agent trained on a given domain size to another domain size: as long as the underlying physics are similar, the policy can be applied to a domain of different dimensions by adapting the number of MARL pseudo-environments deployed. In this section, we perform a deterministic run of the agent trained on the $4\pi$-width and -length domain at \raa{} on a larger domain of size $8\pi$. The number of quadrature points for the CFD solver is doubled in the horizontal directions from $32\times 16 \times 32$ to $64\times 16\times 64$, and the number of control segments is also doubled in each horizontal direction from $8\times 8$ control segments in the $4\pi$ domain to $16\times 16$ control segments in the $8\pi$ domain. In this way, the size of the input state to each agent is kept constant and each agent observes the same segment size as before, \textit{i.e.}, a segment size of $\pi/2 \times \pi/2$. \par 

\begin{figure}
    \centering
    \raisebox{0.25\height}{\includegraphics[width=0.5\textwidth]{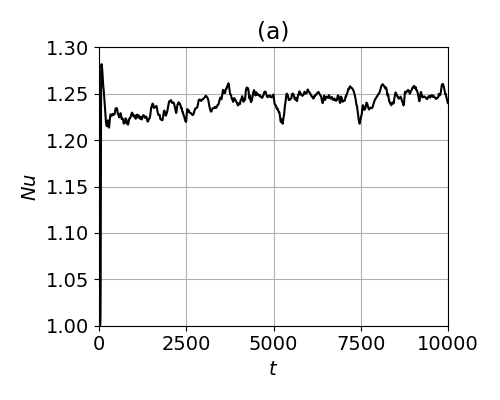}}
    \includegraphics[width=0.45\textwidth]{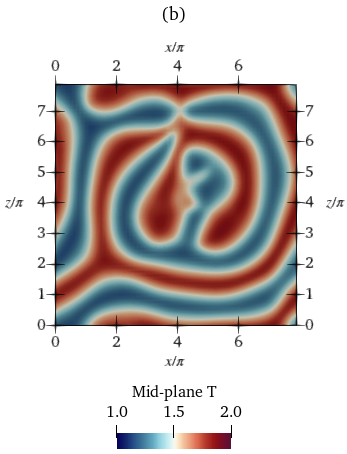}
    \caption{(a) \Nu{} evolution during the baseline simulation in a domain of size $8\pi$ at \raa{} and (b) corresponding mid-plane temperature field at $t=10000$ (right). Video of evolution shown in Online Resource 8.}
    \label{fig:Ra500_8pi_baselineField}
\end{figure}

\begin{figure}
    \centering
    \includegraphics[width=0.65\textwidth]{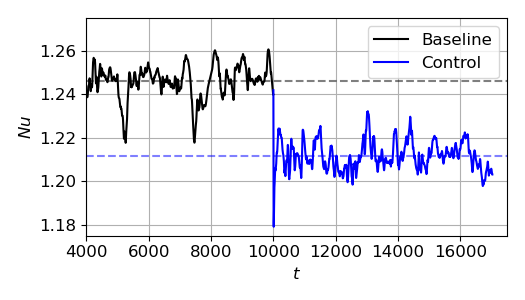}
    \caption{Evolution of \Nu{} in the $8\pi$ domain upon control at \raa{}. The controller used is the agent trained on the $4\pi$ domain at \raa{}. The horizontal broken lines represent the mean \Nu{}, averaged over the last half of the duration of the baseline, \Nuref{},  (black) and over the duration of the control (blue).}
    \label{fig:Ra500_8pi_control_nu}
\end{figure}

\begin{figure}
    \centering
    \includegraphics[width=0.9\textwidth]{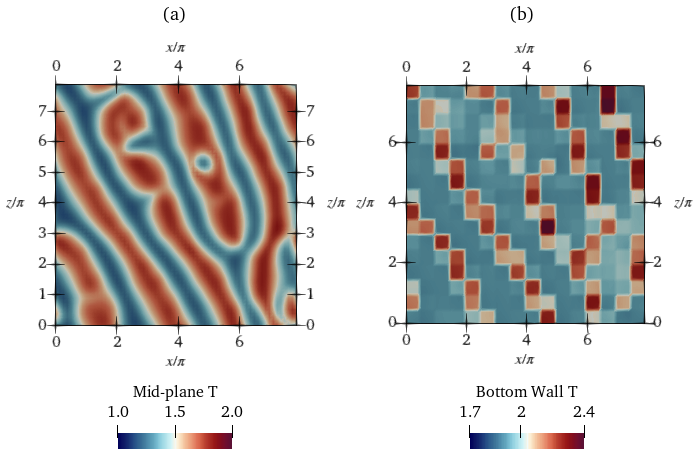}
    \caption{The temperature field at the mid-plane (a), and control actuations at the bottom wall of the $8\pi$ domain (b), obtained from the control executed by the agent trained on the $4\pi$ domain. The time of the temperature field corresponds to the final time step in Fig. \ref{fig:Ra500_8pi_control_nu}. Video of control shown in Online Resource 9.}
    \label{fig:Ra500_8pi_control_field}
\end{figure}

We first show the \Nu{} and temperature field of the baseline state in the mid-plane of the $8\pi$ domain in Fig. \ref{fig:Ra500_8pi_baselineField}. Several curved convection rolls are observed, and the \Nu{} stabilises at a \Nuref{} (time-averaged over the last half of the baseline time, $5000$ to $10000$) of $1.246$. In Fig. \ref{fig:Ra500_8pi_control_nu} we show the evolution of the \Nu{} of the baseline state and controlled states. By applying the $4\pi$ domain-trained agent, the \Nu{} reduces and reaches a mean value of $\Nu_C=1.211$ in the controlled state. Here, $\Nu_C$ is computed from $t=11500$ to the end of the simulation. Using Eq. \eqref{eq:nusselt_percent_reduction_deterministic}, this is a \Nu{} reduction of $14.2\%$. Next, we show the flow topology at the mid-plane of the final controlled state in Fig. \ref{fig:Ra500_8pi_control_field}. From the plot, it is visible that the baseline convection rolls are broken and reconfigured to a shape with diagonally running convection rolls that are nearly straight. This pattern is qualitatively similar to the end control states reached by the deterministic evaluation of the agent in \raa{} and \rab{} in figures \ref{fig:Ra500det_field}(b) and \ref{fig:Ra750det_fields}(b), respectively, and is topologically closer to the configuration observed at lower \Ra{} (Fig. \ref{fig:Ra400}). \par 

\section{Conclusion}\label{sec:conclusion}

In the present study, we apply control using deep reinforcement learning (DRL) on a 3D Rayleigh--Bénard convection (RBC) system, as an extension of the work in 2D by \citet{vignon2023marlrbc}. We use the multi-agent reinforcement learning (MARL) framework that involves several agents, each with its own pseudo-environment. Each agent has its own stream of state observations and rewards from its corresponding local portion of the domain, and applies an independent control action. MARL is especially beneficial for physical systems such as RBC whose dynamics exhibit spatial translational invariance, and allows to overcome the curse of dimensionality, which is otherwise making trial-and-error learning of distributed output control laws computationally intractable. Compared to the 2D case in \citet{vignon2023marlrbc}, the larger control dimensionality (number of actuators required), the larger state space dimensionality and the more complex flow dynamics in the 3D case motivate the need for MARL-based control in 3D. This is, to the best of our knowledge, the first time that DRL control is applied to the 3D RBC system. \par 

The RBC computational domain size is $4\pi \times 2 \times 4\pi$, and the bottom wall is divided into $8\times 8$ square segments of size $0.5\pi \times 0.5\pi$, each of which is assigned to a single MARL agent that applies a temperature value across the segment as its control actuation. Training is performed at \raa{} and \rab{}.  With MARL, control of the RBC system is successfully achieved. During training, the training curves show a drop in \Nu{} by $22\%$ for \raa{} and $12.7\%$ for \rab{}. Both \raa{} and \rab{} are trained with each episode starting from the end state of the baseline for $80\%$ of the total episodes and the remaining $20\%$ starts from the end of the previous episode. In this way, the system also explores trajectories spanning a longer time across successive episodes durations, and thus the agent observes more states, including states representative of the long-term behavior of the system. The case of \rab{} uses three different classes of baselines with different convective patterns, and hence the trained agent is more robust than if trained using a single baseline. During deterministic evaluation, the agent trained at \raa{} achieves $23.5\%$ reduction in \Nu{}, and the agent trained at \rab{} achieves $17.7\%$, $19.3\%$ and $8.7\%$ \Nu{} reduction on baselines of classes 1, 2 and 3, respectively. The fact that the agent is able to control all three classes of baselines at \rab{} is an indicator of its robustness. The structure of the convection in the domain is modified upon control from a spatially irregular pattern to straight convection rolls that run diagonally across the domain. The structure of the convection rolls in the controlled state is topologically similar to that of a RBC system at a lower $\Ra=400$. \par

Comparisons have also been made with proportional control (PC) using a range of coefficients of proportionality $K_p$. Results show that the DRL-based control outperforms PC. At \raa{}, PC achieves a $19.8\%$ \Nu{} reduction and at \rab{}, PC achieves a $6.8\%$ reduction, while DRL-based control achieves $23.5\%$ and $8.7\%$, respectively. Using DRL-based control, in a single training run, the agent learns a more complex control policy than PC, which includes actuator delays and magnitudes of control actions that are different across different segments. As the MARL strategy is adopted in this study, we also show that an agent trained on the $4\pi$ domain can be extended in application to control the RBC system at \raa{} in a domain of a larger size, \textit{i.e.}, $8\pi$-domain. Extending control to larger domains highlights the advantage that MARL can exploit translational invariances of the RBC system at \raa{}. \par

These results open opportunities for a number of future investigations. Various configurations of agents can be adopted for the control of RBC at larger domains and larger \Ra{}. For example, instead of a single agent having as input the states corresponding to a single segment, a set of multiple neighbouring state segments can be assigned as the input state to a single agent. The input states would then contain information from a larger window of the domain that is influenced by the action imposed in the control segment. We are also working on investigating how the results presented here can be extended to higher \Ra{} cases, in which convection is both stronger and more chaotic. In other flow-control problems, DRL-based control has shown promising results in the turbulent regimes, such as 3D cylinders \cite{suarez2024active}, so one could also expect positive results for higher \Ra{} number cases. \par 

 DRL-based control has shown promising results in the turbulent regimes for a 3D cylinder, as we see in e.g. flow control for the wake of 3D cylinder (Suarez et al. 2024)

We believe that the present work is an important milestone that takes DRL from simple 2D cases \citep{vignon2023marlrbc} to complex non-stationary 3D cases. In the future, we will push this further to consider fully turbulent 3D cases, which are relevant for many industrial applications, such as chemical reactors, material manufacturing, and thermal energy systems, where maintaining stable temperature gradients and flow patterns is crucial in order to prevent undesirable instabilities and to enhance efficiency. In other examples such as thermal insulation of buildings, understanding the principles of control in RBC can be crucial to minimize the energy losses of the heating or cooling system.

\backmatter

\bmhead{Acknowledgements} 
The authors extend their acknowledgement to the National Academic Infrastructure for Supercomputing in Sweden (NAISS) for resources provided for this study, and to the European Research Council (ERC) for financial support. Views and opinions expressed are those of the authors only and do not necessarily reflect those of the European Union or the European Research Council. Neither the European Union nor the granting authority can be held responsible for them. 

\section*{Declarations}

\bmhead{Funding Information}
Part of the deep-learning-model training was enabled by resources provided by the National Academic Infrastructure for Supercomputing in Sweden (NAISS). Ricardo Vinuesa acknowledges financial support from ERC grant no. `2021-CoG-101043998, DEEPCONTROL'. 

\bmhead{Conflict of Interest} 
The authors declare no conflict of interest.

\bmhead{Author Contributions} 
Conceptualization: Ricardo Vinuesa, Jean Rabault, Joel Vasanth, Francisco Alcántara-Ávila; Methodology: Ricardo Vinuesa, Jean Rabault, Joel Vasanth, Francisco Alcántara-Ávila; Formal analysis and investigation: Joel Vasanth; Software: Joel Vasanth, Francisco Alcántara-Ávila and Mikael Mortensen;  Writing - original draft preparation: Joel Vasanth; Writing - review and editing: Joel Vasanth, Jean Rabault, Francisco Alcántara-Ávila, Mikael Mortensen, Ricardo Vinuesa; Funding acquisition: Ricardo Vinuesa; Resources: Ricardo Vinuesa; Supervision: Jean Rabault, Ricardo Vinuesa, Francisco Alcántara-Ávila. All authors read and approved the final manuscript.

\bmhead{Data Availability}
The data that support the findings of this study are available from the corresponding author upon reasonable request.

\begin{appendices}

\section{Convergence Analysis}\label{sec:appendix_convergence}
\begin{figure}[h!]
\centering
\includegraphics[width=1.0\textwidth]{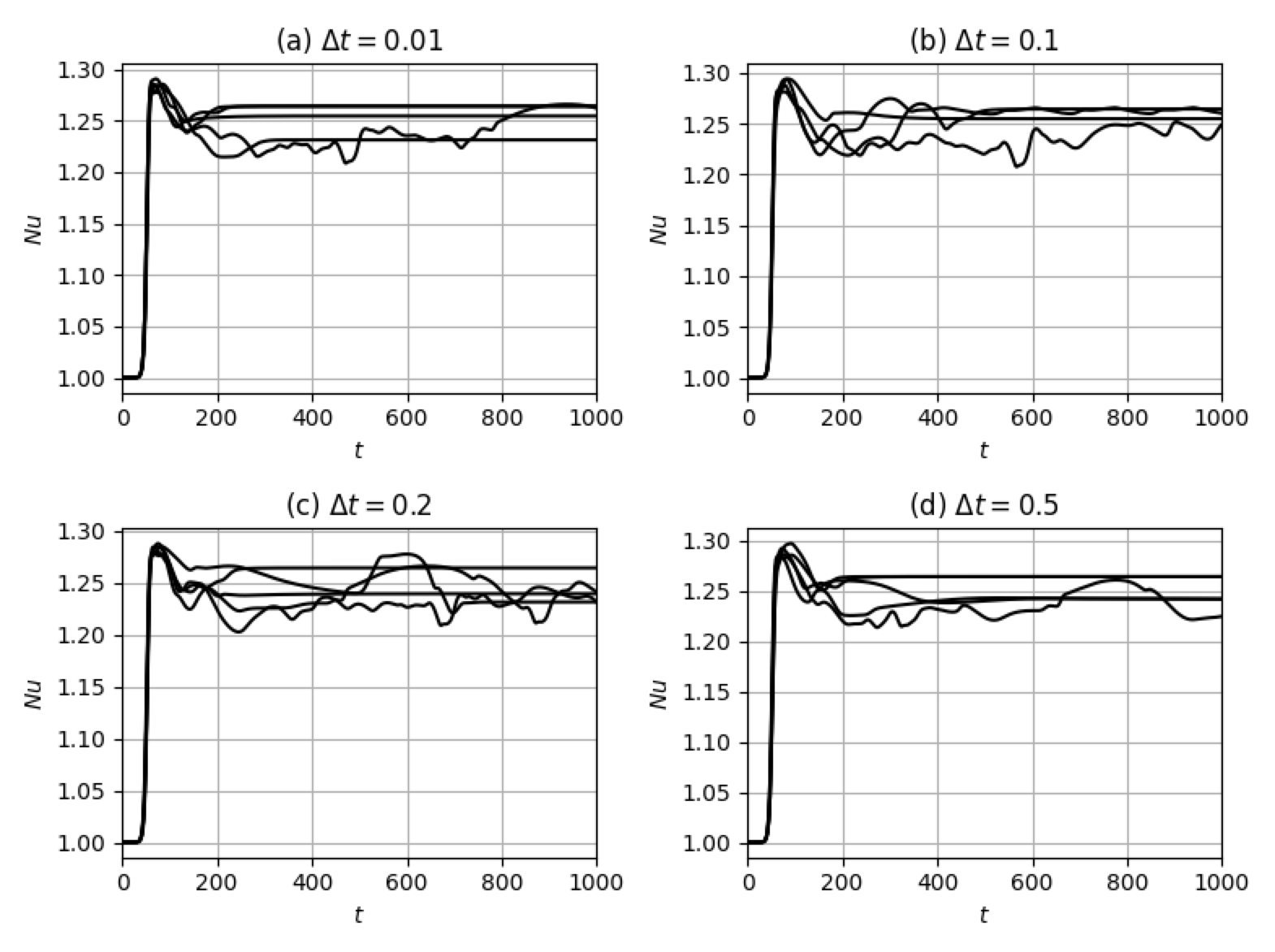}
\caption{Time evolutions of Nu with various time steps $\mathrm{\Delta t}$. For each $\mathrm{\Delta t}$, five simulations with the same parameters are shown.}
\label{fig:appendixA-baselines_dt}
\end{figure}

In this section, we simulate the baseline for $Ra=500$ with different time steps $\mathrm{\Delta t}=0.01,\ 0.1,\  0.2$ and $0.5$ in order to justify our choice of $\mathrm{\Delta t}=0.1$. For each $\mathrm{\Delta t}$, we run five simulations with the same parameters. We note here that for $\mathrm{\Delta t} > 0.5$, the solution is found to be numerically unstable and the values of Nu obtained are Nan. Shown in Fig. \ref{fig:appendixA-baselines_dt} are the resulting Nu evolutions for $\mathrm{\Delta t} \leq 0.5$. We see that in each case, the Nu can reach either a steady or unsteady solution with small fluctuations about the mean. We also note that the mean Nu (averaged from $t=400-1000$) for each of the five simulations for each $\mathrm{\Delta t}$, lies within the range of 1.23-1.27. \par
\begin{figure}[h!]
\centering
\includegraphics[width=0.75\textwidth]{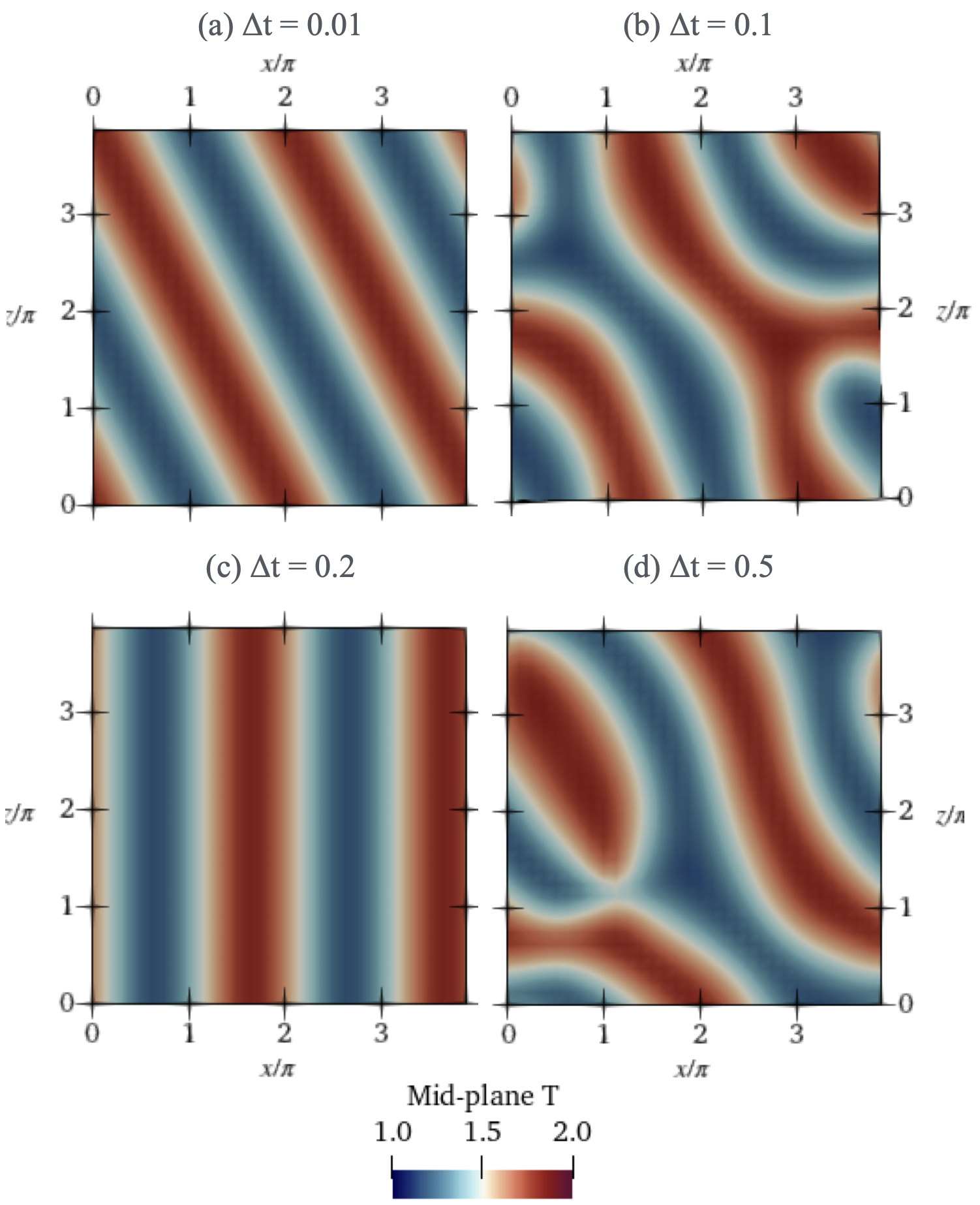}
\caption{Mid-plane temperature fields at $t=1000$ for the simulations with different time steps as shown in Fig. \ref{fig:appendixA-baselines_dt}.}
\label{fig:appendixA-temps_dt}
\end{figure}

 
We also observe the flow structures in the temperature fields shown in Fig. \ref{fig:appendixA-temps_dt}. The temperature fields for each $\mathrm{\Delta t}$ are from a single randomly selected simulation out of the five solutions, at the end time $t=1000$. By visual inspection, we note that the convection rolls have a length scale of $\pi$ that is roughly uniform along the axis of the rolls. Thus, solutions with $\mathrm{\Delta t}
\leq 0.5$ qualitatively have the same spatial and temporal features, and we deem any of them fit to use as baselines for control. \par 

The thermal relaxation time for the non-dimensional Rayleigh-Bénard problem is $\sqrt{\mathrm{Ra}\cdot\mathrm{Pr}}$ \citep{bergedubois1984} which for $Ra = 500$ and $\mathrm{Pr} = 0.7$ is $\approx 18.7$ non-dimensional time units. A motivating factor for selecting $\mathrm{\Delta t}=0.1$ (instead of 0.01 or any value $\leq 0.5$) for the remainder of the paper is that this is small enough to be at least 1/100th the thermal relaxation time, and large enough to ensure quicker simulations. Furthermore, during the control simulations, a low $\mathrm{\Delta t}$ ensures that any numerical instabilities that may arise due to the sharp changes in temperature actuations at the bottom wall can be prevented.

\section{Supplementary information}\label{sec:appendix_supplementary_information}

Several supplementary files are provided along with this article. These are video animations of temperature-field evolutions from simulation results corresponding to various figures in the main text. An exhaustive list of the video filenames with their captions is provided below. The reader is advised to refer to this list when `Online Resource' is mentioned in the main text. Note that `Online Resource' is abbreviated as `OR' below. \par 

\begin{enumerate}
    \item OR 1: `\verb|or1_fig4a_Ra500_baseline.avi|'. Mid-plane temperature field evolution of the baseline of \raa{}. Corresponding figure - Fig. \ref{fig:BaselinesNu_topology}(a).
    \item OR 2: `\verb|or2_fig4b_Ra750_baseline.avi|'. Mid-plane temperature field evolution of the class 3 baseline of \rab{} (run 5 in Fig. \ref{fig:BaselinesNu_topology}(b)).
    \item OR 3: `\verb|or3_fig9_Ra500_det.avi|'. Mid-plane and bottom wall temperature fields during the evolution of the RBC system under action of the MARL controller. At time $1300$, a more organised regular spatial arrangement of convection rolls is observed, characterised by straight diagonal running cells. Corresponding figure - Fig. \ref{fig:Ra500det_field}.
    \item OR 4: `\verb|or4_fig10_Ra400_baseline.avi|'. Evolution of the baseline at $\mathrm{Ra}=400$. Time units shown correspond to those in Fig. \ref{fig:Ra400}(a).
    \item OR 5: `\verb|or5_fig12_Ra750_det.avi|'. Deterministic evaluation of agent trained at \rab{}. Irregular baseline shape broken down and convection structure becomes regular, \textit{i.e.}, straight diagonal cells at time $5928$. Corresponding figure - Fig. \ref{fig:Ra750det_fields}
    \item OR 6: `\verb|or6_fig15_Ra500_prop_control.avi|'. Proportional control for \raa{} at $K_p=0.6$.  Corresponding figure - Fig. \ref{fig:Ra500prop_field}.
    \item OR 7: `\verb|or7_fig16_Ra750_prop_control.avi|'. Proportional control for \rab{} at $K_p=1.1$.  Corresponding figure - Fig. \ref{fig:Ra750prop_field}.
    \item OR 8: `\verb|or8_fig17_Ra500_8pi_baseline.avi|'. Baseline evolution for \raa{} in a domain of length $8\pi$. Corresponding figure - Fig. \ref{fig:Ra500_8pi_baselineField}.
    \item OR 9: `\verb|or9_fig19_Ra500_det_8pi_control.avi|'. MARL agent trained on domain of length $4\pi$ used in deterministic evaluation mode for the control of RBC in larger domain of length $8\pi$. Corresponding figure - Fig. \ref{fig:Ra500_8pi_control_field}.
\end{enumerate}

\section{Open source code release on Github}\label{sec:appendix_code_release}
All the codes, scripts, and post-processing tools used in this work will be made available on Github together with readmes and user instructions, over publication of this manuscript. Reasonable user support will be provided through the issue tracker of the corresponding Github repository. The url for the repository is \url{https://github.com/KTH-FlowAI/DRL_MARL_RayleighBenard3D_Control}. \par 
The shenfun CFD case setup is described in great details, including all the numerical implementation considerations, elements chosen, and code walk-through, on the shenfun RBC documentation page: \url{https://shenfun.readthedocs.io/en/latest/rayleighbenard.html}. Note that the conventions used in the present paper and in this documentation page are slightly different: while the present paper uses $x$ as the horizontal (``wall parallel") description and $y$ as the vertical (``wall normal") direction, the documentation page uses the opposite conventions. This has no influence on the CFD results \textit{per se}. The code uses the same conventions as the shenfun documentation, and only the present paper uses separate conventions, for simplicity of the writing and conformity with the literature.

\newpage
\section{Convection structure for \rab{} baselines}\label{sec:appendix_convection_structure}

In this appendix, we show the temperature fields for the baselines for \rab{} (not shown in Fig. \ref{fig:BaselinesNu_topology}(b)) as described in Sec. \ref{subsec:results_baselines}.

\begin{figure}[h]
\centering
\includegraphics[width=0.4\textwidth]{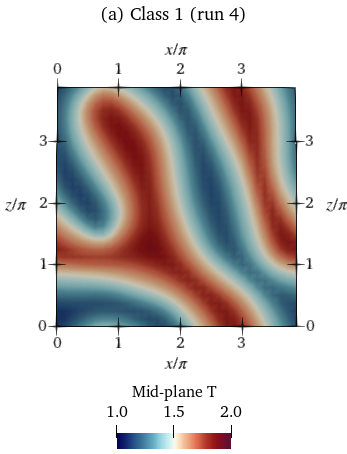}
\includegraphics[width=0.4\textwidth]{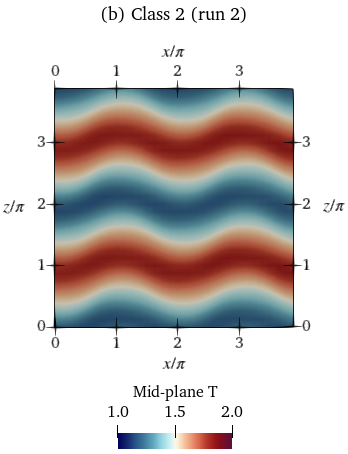}
\includegraphics[width=0.4\textwidth]{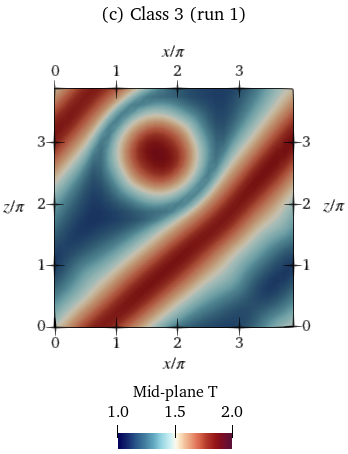}
\includegraphics[width=0.4\textwidth]{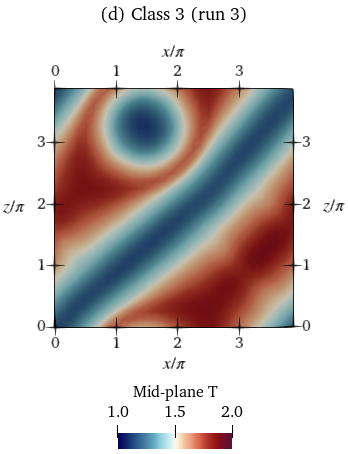}
\caption{Temperature fields of the baselines shown in the mid-plane cross-section ($y=0$) of the domain for \rab{} for (a) Class 1 (b) Class 2. (c-d) runs 1 and 3 of class 3 (Fig. \ref{fig:BaselinesNu}). Run 5 of class 3 is shown in Fig. ~\ref{fig:BaselinesNu_topology}(b).}
\label{fig:appendixA-ra750baselines}
\end{figure}

\end{appendices}



\end{document}